\documentclass[12pt,linenumbers, preprint]{aastex}

\usepackage{amsmath,longtable,natbib}
\usepackage{float}
\usepackage{hyperref}
\usepackage{adjustbox}

\usepackage{caption}
\usepackage{graphicx,epstopdf}
\DeclareGraphicsExtensions{.ps}
\shorttitle{Wide-band timing of GMRT MSPs}
\shortauthors{Sharma et al.}
\usepackage{subfig}
\begin{document}
\captionsetup[figure]{labelfont={bf},labelformat={simple},labelsep=period,name={Figure}}
\captionsetup[table]{labelfont={bf},labelformat={simple},labelsep=period,name={Table}}
\title{Wide-band timing of GMRT discovered millisecond pulsars}
\author{
Shyam~S.~Sharma\altaffilmark{1}, 
Jayanta Roy\altaffilmark{1},
Bhaswati Bhattacharyya\altaffilmark{1},
Lina Levin\altaffilmark{2},
Ben~W.~Stappers\altaffilmark{2},
Timothy T. Pennucci\altaffilmark{3},
Levi Schult\altaffilmark{4},
Shubham Singh\altaffilmark{1},
Aswathy Kaninghat\altaffilmark{1,5}}
\altaffiltext{1}{National Centre for Radio Astrophysics, Tata Institute of Fundamental Research, Pune 411007, India}
\altaffiltext{2}{Jodrell Bank Centre for Astrophysics, School of Physics and Astronomy, The University of Manchester, Manchester M13 9PL, UK}
\altaffiltext{3}{Institute of Physics, Eötvös Loránd University, Pázmány P.s. 1/A, 1117 Budapest, Hungary}
\altaffiltext{4}{Department of Physics and Astronomy, Vanderbilt University, 2301 Vanderbilt Place, Nashville, TN 37235, USA}
\altaffiltext{5}{Leibniz Universität Hannover, Germany}
\affil{}
\section*{ABSTRACT}

Modeling of frequency-dependent effects, contributed by the turbulence in the free electron density of interstellar plasma, is required to enable the detection of the expected imprints from the stochastic gravitational-wave (GW) background in pulsar timing data. In this work, we present an investigation of temporal variations of interstellar medium for a set of millisecond pulsars (MSPs) with the upgraded GMRT aided by large fractional bandwidth at lower observing frequencies. Contrary to the conventional narrow-band analysis using a frequency invariant template profile, we applied \textit{PulsePortraiture}
based wide-band timing analysis while correcting for the evolution of the pulsar profile with frequency. 
Implementation of \textit{PulsePortraiture} based wide-band timing method for the GMRT discovered MSPs to probe the DM variations resulted in a DM precision of $10^{-4}\,pc~cm^{-3}$. In general, we achieve {similar} DM and timing precision from wide-band timing compared to the narrow-band timing with matching temporal variations of DMs. This wide-band timing study of newly discovered MSPs over a wide frequency range highlights the effectiveness of profile-modeling at low frequencies and probes the potential of using them in pulsar timing array.

\section{Introduction}
\label{sec:intro}
Millisecond pulsars (MSPs) are fast rotating neutron stars with exceptional rotational stability enabling the precise determination of their rotational and orbital (for systems in binary) properties as well as using them to probe the interstellar medium (ISM) ({e.g., \cite{1990ApJ...364..123F}}). {Such exceptional stability of MSPs also allows them to use as a probe to search for gravitational waves (GWs).}

The stochastic GWs background manifests as an unmodeled effect in the timing residuals {(known as timing noise)} whose detectability depends on the timing span and precision of the measurements \citep{2013CQGra..30v4015S}. The Pulsar Timing Array (PTA) experiment {(e.g., \cite{1979ApJ...234.1100D})} uses a set of MSPs with different angular separations in the sky to search for the angular correlation between the residuals of the arrival times of pairs of pulsars { \citep{1983ApJ...265L..39H}}. Such correlation reveals the signature of low-frequency stochastic isotropic GW background in the timing data, where the largest contribution is thought to be coming from an ensemble of merging super-massive black hole binaries \citep{2019A&ARv..27....5B}. 

One of the crucial challenges for PTAs is to disentangle and mitigate the timing noise contributed by variations in the free electron density of the interstellar plasma. Time-varying ISM effects (i.e., changes in the dispersion measure, the influence of scattering) on pulse arrival time need to be precisely determined to improve the timing precision. 
The emission from the pulsar undergoes frequency dependent effects as it propagates through the ISM. A signal of frequency $\nu$ arrives at Earth at a delayed time $\Delta t_{\nu}$, with respect to infinite frequency, which is given by
\begin{equation}
    \Delta t_{\nu}=K\times DM\,\nu^{-2}
\end{equation}
where {K is the dispersion constant} with value of 4.148808(3) GHz$^2\,$cm$^3\,$pc$^{-1}\,$ms and DM (dispersion measure) {is the free electron column density integrated along the line of sight (LOS) from the observer to the source, i.e.,}
\begin{equation}
    DM\equiv \int_{LOS} n_e dl \quad . 
\end{equation}
Equation (1) shows that a typical DM variation of $10^{-3}-10^{-4}$ $\,pc~cm^{-3}$, seen in pulsar observations {(e.g., \cite{2020A&A...644A.153D})},
 introduces a change in pulse time of arrival (ToA) of more than $1 \mu s$ at $\nu\sim$ 1 GHz (with respect to the infinite frequency). Whereas, to achieve timing precision better than 100 ns at an observing frequency of 1400 MHz, the DM variation needs to be modeled at a precision of $\sim$ $10^{-5}$ $\,pc~cm^{-3}$ \citep{2007MNRAS.378..493Y}.

 {Scattering of radio signals by inhomogeneities in the ISM causes frequency-dependent delays in the ToAs. For a basic model of ISM, assuming a thin screen of plasma located between the pulsar and the observer \citep{1968Natur.218..920S}, scattering delays can be measured using scintillation pattern on the dynamic spectra. The scattering delay $\tau_{scat.}$ is proportional to $\nu^{-4}$.  So at lower frequencies, the DM effect ($\nu^{-2}$) can be more distinguishable from the scattering ($\nu^{-4}$) to reduce the covariance between scattering and DM effects while fitting for DM.   
 Timing experiments at higher frequency usually find scattering delay smaller than the ToA uncertainties implying that the variations of such delay are not having much adverse effect in the timing precision (e.g., \cite{2016ApJ...818..166L}, \cite{2021ApJ...917...10T}).}
 
Data collected by the International Pulsar Timing Array (IPTA) consists of several MSPs ($\sim65$, \cite{2019MNRAS.490.4666P}) observed over a wide range of frequencies ($0.3-3.1$ GHz) with various telescopes. {It aims to improve the PTA sensitivity to GW signals by combining data from the three individual PTAs [North American Nanohertz Observatory for Gravitational Waves (NANOGrav) \citep{2009arXiv0909.1058J}; European Pulsar Timing Array (EPTA) \citep{2006ChJAS...6b.298S}; Parkes pulsar timing array (PPTA) \citep{2006ChJAS...6b.139M}].} {Due to the greater severity of ISM effects at low frequencies}, frequencies greater than 1 GHz are preferred for high-precision timing analysis. However, low-frequencies (i.e., $<$ 1 GHz) can provide a sensitive probe for measuring DM and its temporal evolution to mitigate adverse effects in the arrival times (\cite{2012A&A...543A..66H}) which are embedded in the high frequency measurements.

{The Giant Metrewave Radio Telescope (GMRT)} is one of the most sensitive radio telescopes at low radio frequencies and covering a frequency range from 120 to 1460 MHz. The GMRT being a IPTA telescope can provide sensitive low-frequency timing measurements, which are already demonstrated by \cite{2020arXiv200908409J} with the legacy GMRT {\citep{2010ExA....28...25R}, \cite{2021arXiv210105334K} and \cite{2021arXiv211206908N} with the upgraded GMRT (uGMRT; \cite{2017JAI.....641011R}, \cite{2017CSci..113..707G}). 
The current observing setup for the observations presented in this paper aims to utilise the maximum possible sensitivity at the low frequencies with the GMRT and it is different from the regular Indian PTA (InPTA) monitoring program.} 
Band-3 of the uGMRT (i.e. 300 to 500 MHz) with its large fractional bandwidth provides a facility for very accurate intra-band DM estimates. {Precise DM measurements obtained from this band can be used to correct for dispersive delays in simultaneous high-frequency timing data. {However, due to multi-path scattering of pulsar signals the DM can be different at lower and higher frequencies as shown by \cite{2016ApJ...817...16C}.}}  According to the radiometer equation \citep{2012hpa..book.....L}, a larger observing bandwidth results in a higher signal-to-noise ratio (S/N) pulse profile promising better ToA and DM precision. However, the intrinsic pulse profile can evolve significantly with frequency within an observing band. In addition, at lower frequency {band of uGMRT with larger fractional bandwidth}, radio frequency interference (RFI), scintillation, and scattering can contaminate the pulse detection significance.

The standard narrow-band (NB) timing technique {\citep{2021ApJS..252....4A}} uses a single frequency averaged template to generate ToAs for different subbands within the observing bandwidth. The NB technique doesn't account for any frequency-dependent effects. It works adequately at high frequencies with smaller fractional bandwidth where the frequency-dependent effects {within the band are less compared to lower frequencies with larger fractional bandwidth.} {\cite{2014ApJ...790...93P} and \cite{2014MNRAS.443.3752L}} describes the simultaneous wide-band (WB) ToA and DM measurement technique using a frequency-dependent template.  \cite{2019ApJ...871...34P} developed a principal-component-decomposition based modeling of pulse profiles as a function of frequency, which is an input to the WB ToA and DM measurement technique. All of these are implemented in a package called ``$\textit{PulsePortraiture}$"\footnote{\label{pulse_portraiture}\url{https://github.com/pennucci/PulsePortraiture}} \citep{2016ascl.soft06013P}. Using {\textit{PulsePortraiture}$^{\ref{pulse_portraiture}}$}, we can estimate the ToA and DM simultaneously at a high precision with frequency-dependent template.

{\cite{2021ApJS..252....5A} reported} WB timing results for 47 NANOGrav MSPs with a range of frequency coverage: 1.4 GHz (with a bandwidth of $\sim$ 600 MHz), 800 MHz (with a bandwidth of $\sim$ 186 MHz) and 430 MHz (with a bandwidth 25$-$50 MHz). The detailed comparisons with NB timing results for these MSPs establish the potential of WB timing in achieving higher timing and DM precision. 

{In this work,} we present the results of applying WB timing analysis for four GMRT discovered MSPs with the uGMRT in band-3 (300$-$500 MHz) and band-4 (550$-$750 MHz). {We validate the WB analysis pipeline with a few bright PTA MSPs observed with the uGMRT in band-3 and band-5 (1060$-$1460 MHz)}. Since the frequency-dependent effects are much more prominent at low frequencies, the observing bands of uGMRT, specially band-3 with 0.5 fractional bandwidth, demonstrate effectiveness of pulse-profile modeling with frequency. Observation and data processing details are provided in section \ref{sec:Observations}. Section \ref{sec:techniques} contains the details of the NB and WB timing techniques. Section \ref{sec:Results} contains the measurements obtained from two timing analysis and comparisons with some of the existing results (\cite{2021arXiv210105334K}, \cite{2021ApJS..252....5A} and \cite{2021arXiv211206908N}). In Section \ref{sec:conclusions} we summarise the improvements seen with the WB timing analysis.

\section{Observations and data processing}
\label{sec:Observations}

\begin{figure}[H]
   
    \subfloat[J1640+2224, B3\label{b}]{{\includegraphics[width=0.243\linewidth,height=5cm]{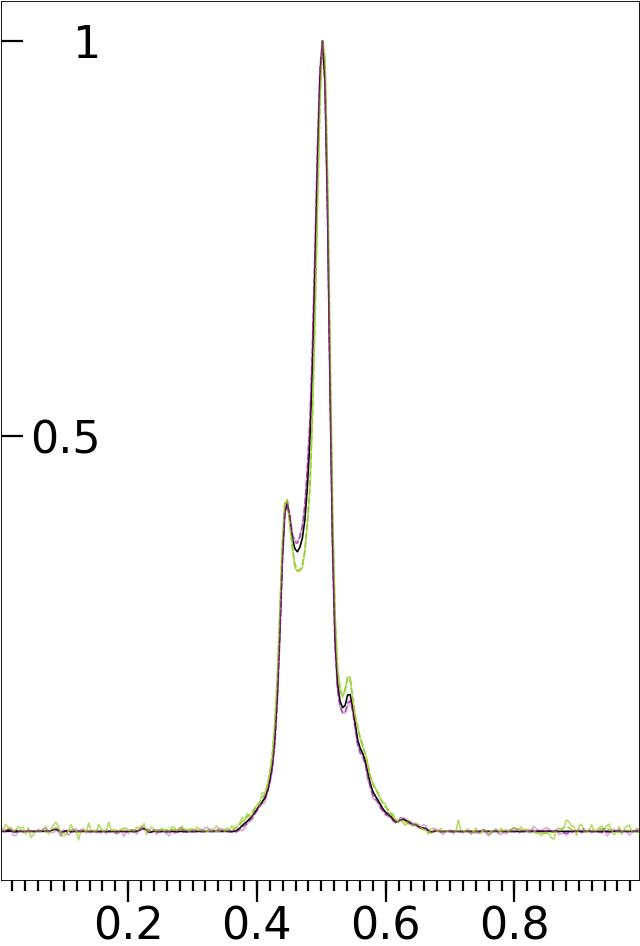} }}
    \subfloat[J1713+0747, B3\label{d}]{{\includegraphics[width=0.243\linewidth,height=5cm]{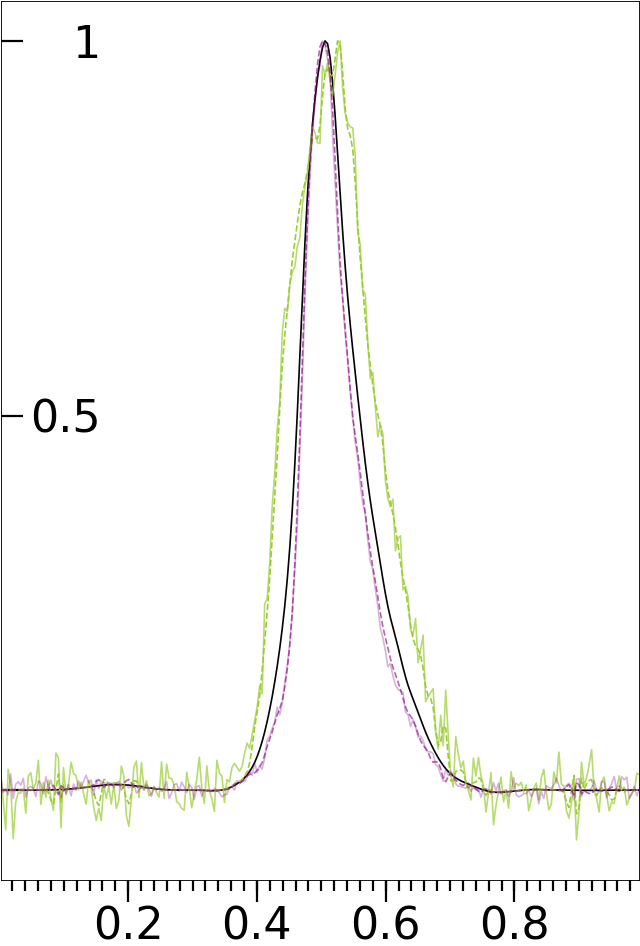} }}
    \subfloat[J1909-3744, B3\label{c}]{{\includegraphics[width=0.243\linewidth,height=5cm]{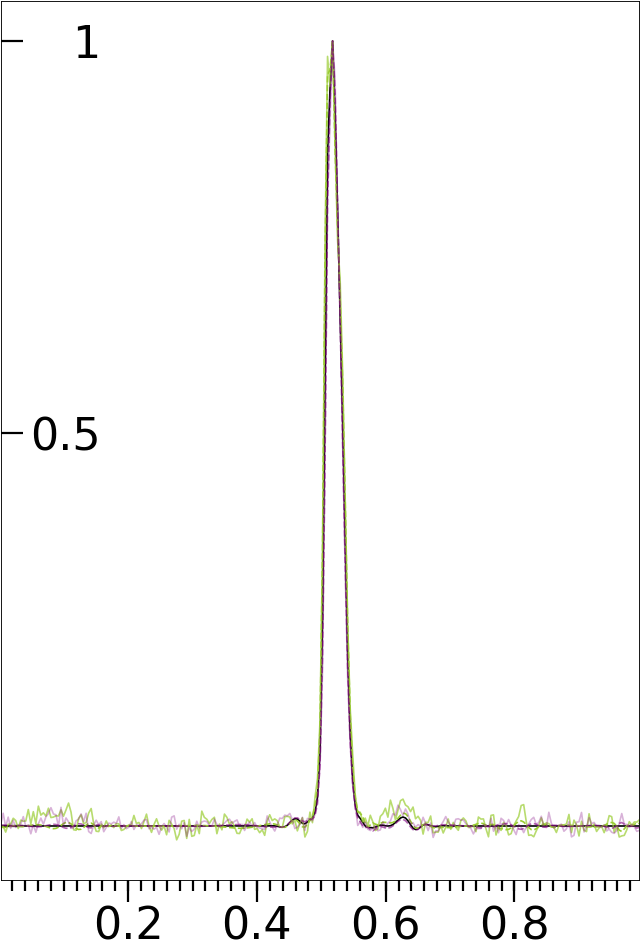} }}
    \subfloat[J2145-0750, B3\label{a}]{{\includegraphics[width=0.243\linewidth,height=5cm]{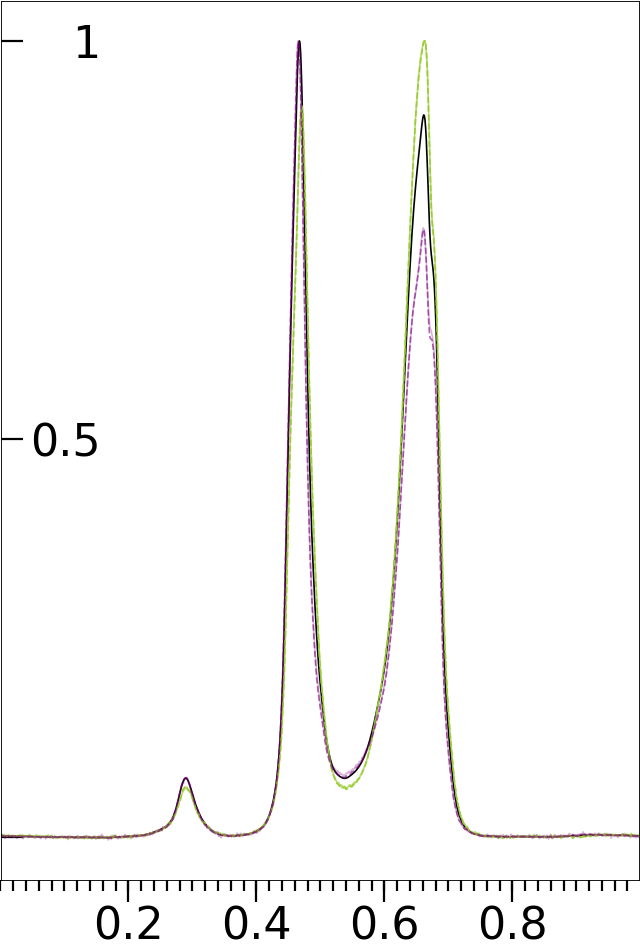} }}

    \subfloat[J1640+2224, B5\label{f}]{{\includegraphics[width=0.243\linewidth,height=5cm]{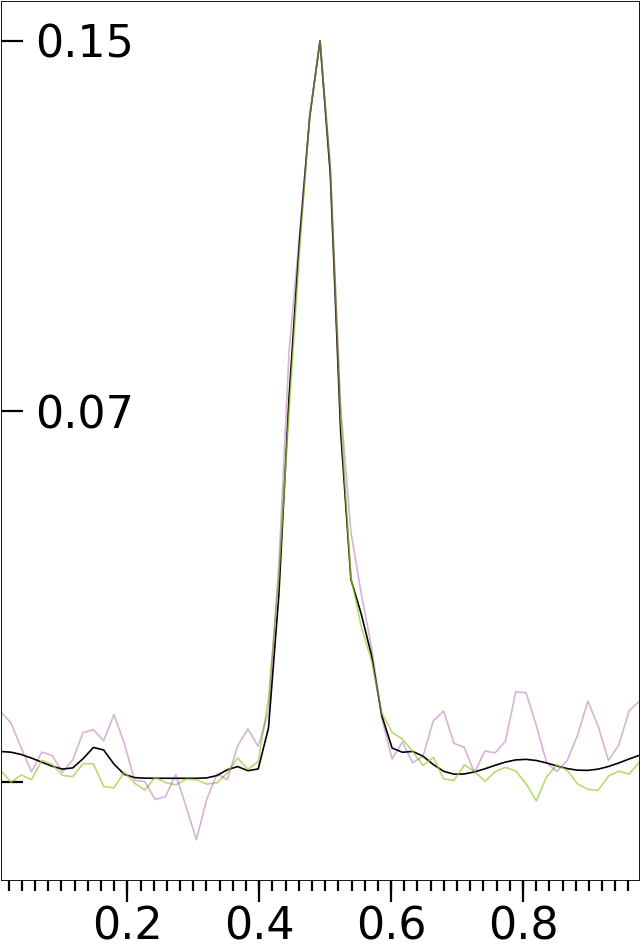} }}
    \subfloat[J1713+0747, B5\label{h}]{{\includegraphics[width=0.243\linewidth,height=5cm]{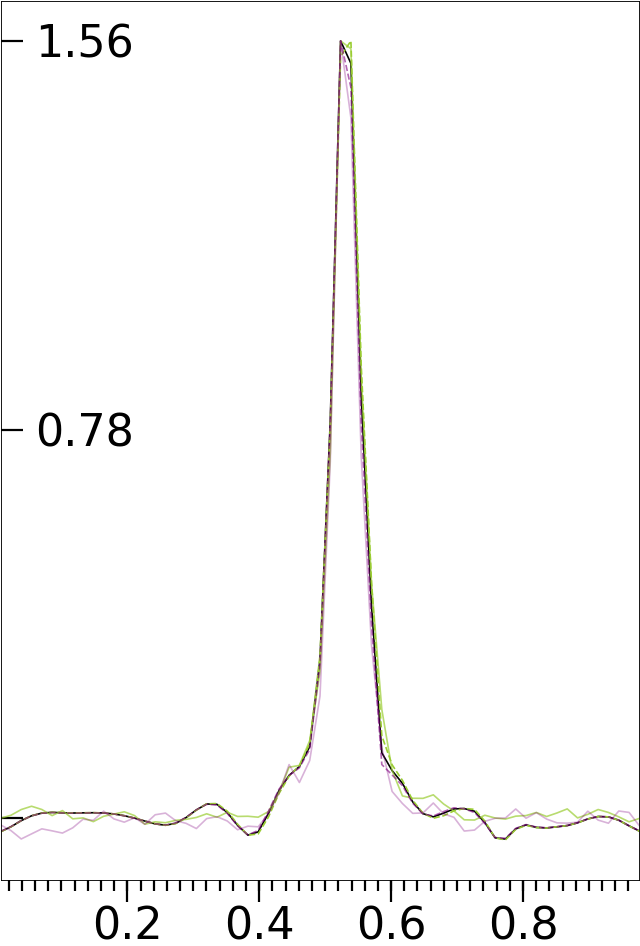} }}
    \subfloat[J1909-3744, B5\label{g}]{{\includegraphics[width=0.243\linewidth,height=5cm]{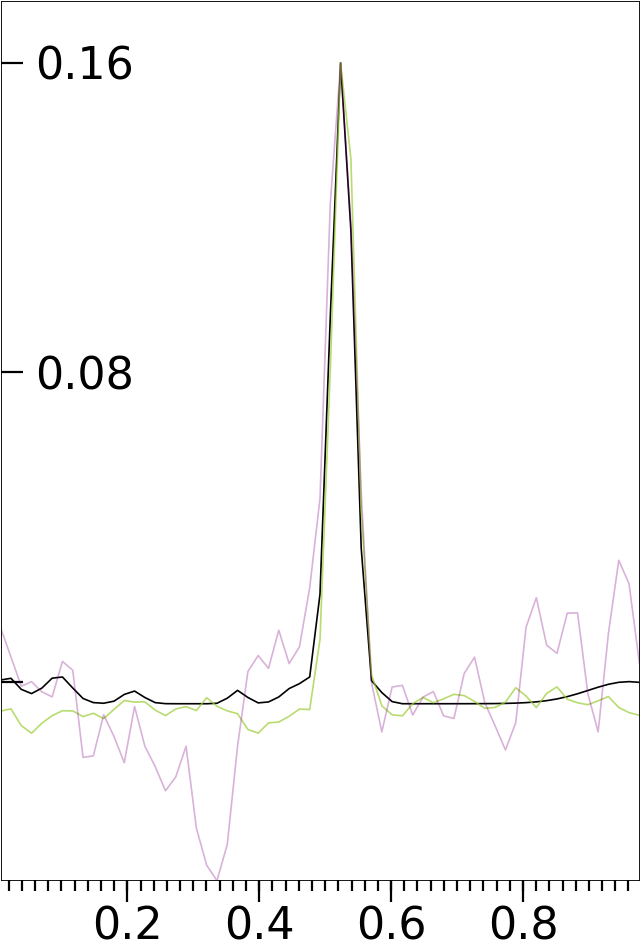} }}
    \subfloat[J2145-0750, B5\label{e}]{{\includegraphics[width=0.243\linewidth,height=5cm]{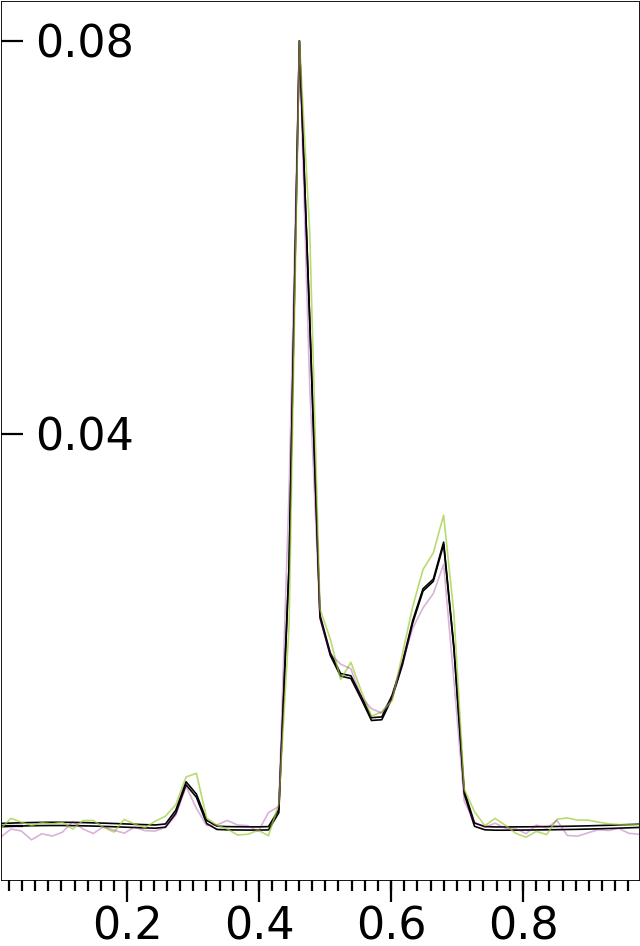} }}
    \caption{{Pulse profiles of the PTA pulsars in band-3 (B3) and band-5 (B5) of the uGMRT. {The profiles in band-3 are normalised by their peak intensity values. In band-5, the normalised profiles are scaled by the ratio of S/N in band-5 to band-3}. We combined all high S/N observations in each of the uGMRT bands separately. We averaged the combined data in time and divided it {into four subbands of equal widths}. To highlight the profile evolution with frequency, we have plotted the profile of pulsar for the {highest-frequency subband} (centered at 460 MHz and 1400 MHz in B3 and B5, respectively) and the {lowest-frequency subband} (centered at 330 MHz and 1110 MHz in B3 and B5, respectively) of a given band. 
The yellow-green solid and dashed curves represent {the lowest-frequency subband} data profile and fitted smooth curve, respectively. 
We have used light-purple solid and dashed curves for {the highest-frequency subband} data and smoothen profile, respectively. The black solid curve represents the NB template averaged over the full band used to extract NB ToAs. Table \ref{table_1} lists the integration times for individual epochs for all pulsars.}}
    \label{PTA_pulsar_profiles}
    \end{figure}

\begin{figure}[H]
    \subfloat[J1120-3618, B3\label{l}]{{\includegraphics[width=0.243\linewidth,height=5cm]{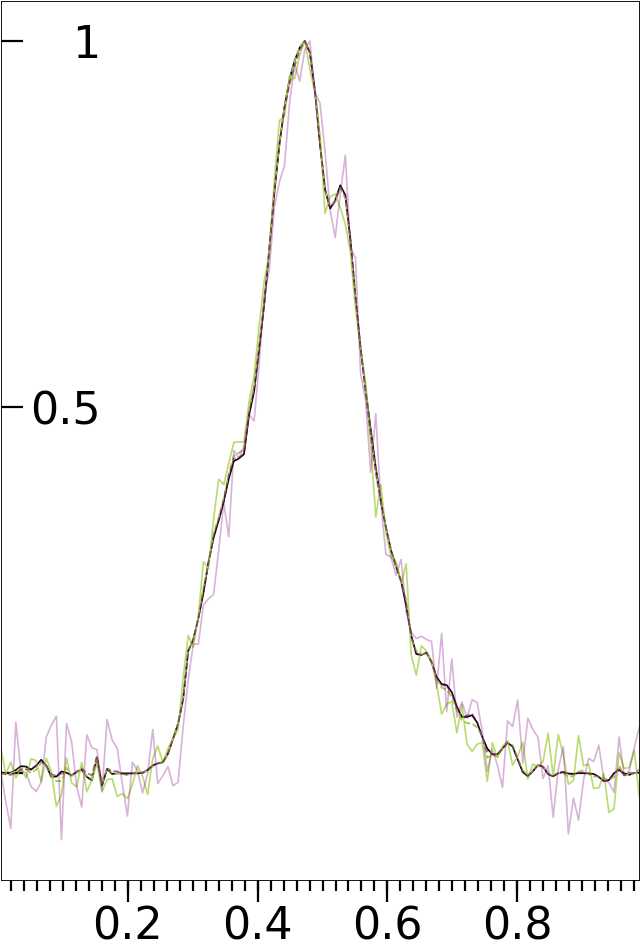} }}
    \subfloat[J1646-2142, B3\label{j}]{{\includegraphics[width=0.243\linewidth,height=5cm]{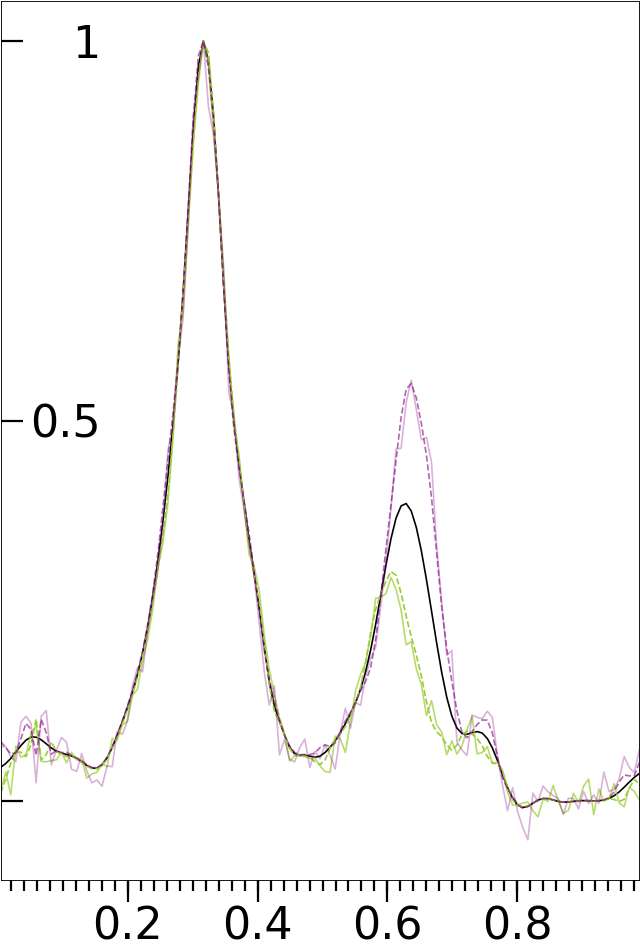} }}
    \subfloat[J1828+0625, B3\label{i}]{{\includegraphics[width=0.243\linewidth,height=5cm]{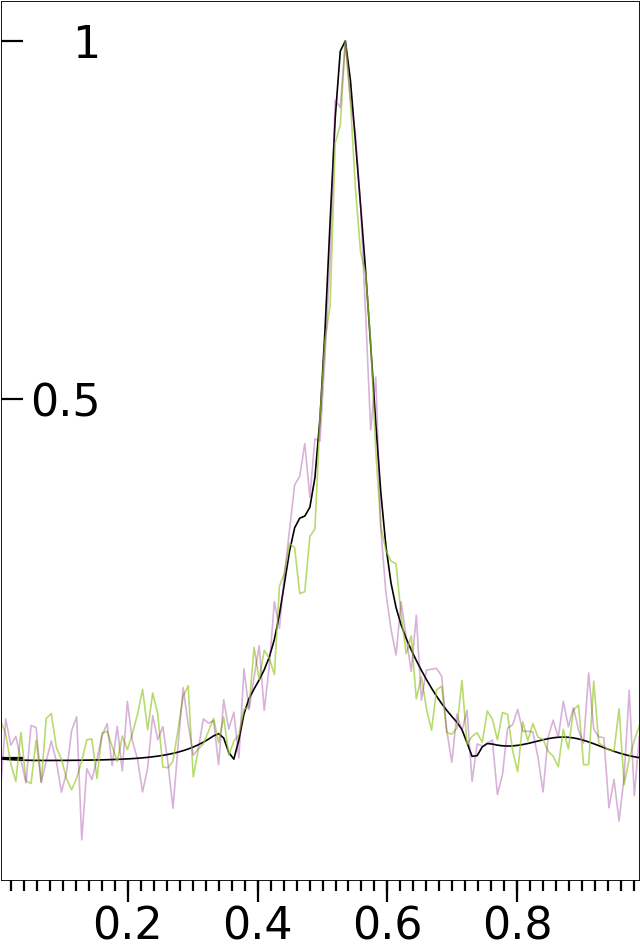} }}
    \subfloat[J2144-5237, B3\label{k}]{{\includegraphics[width=0.243\linewidth,height=5cm]{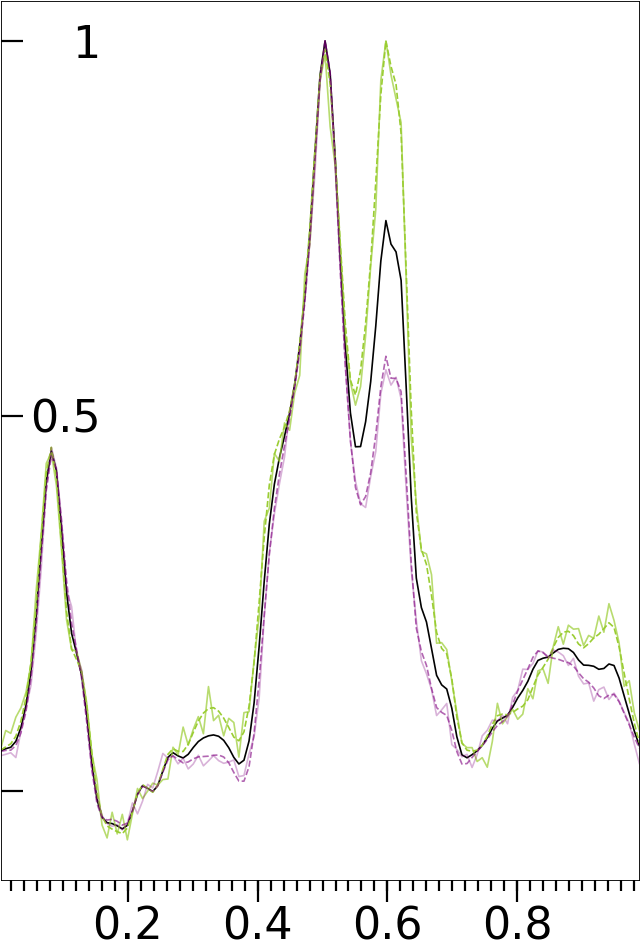} }}
    
    ~~~~~~~~~~~~~~~~~~~~~~~~~~~~~~~~~~~~~~~~~
    \subfloat[J1646-2142, B4\label{m}]{{\includegraphics[width=0.243\linewidth,height=5cm]{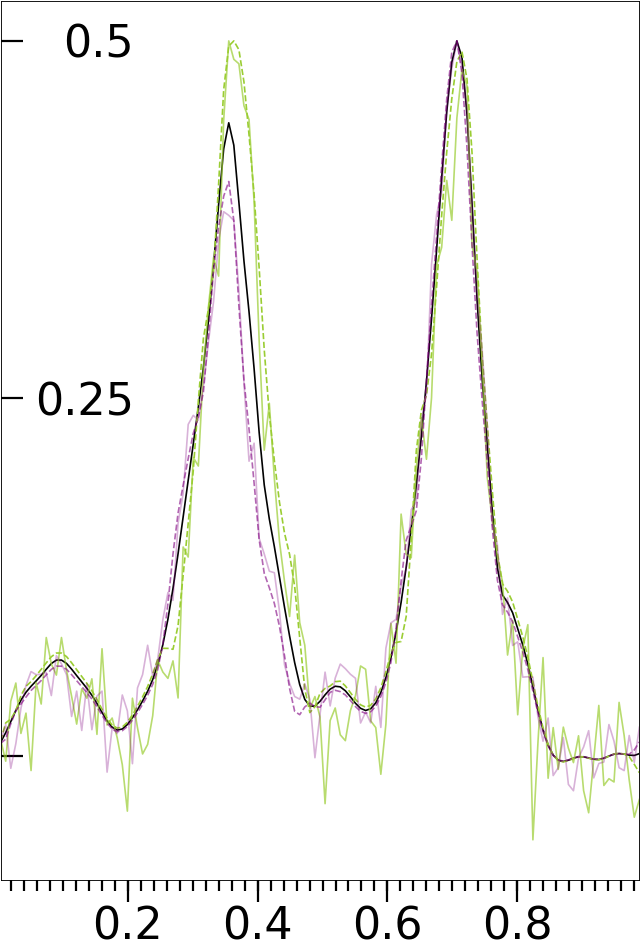} }}
    ~~~~~~~~~~~~~~~~~~~~~~~~~~~~~~~~~~~~~~~~
    \subfloat[J2144-5237, B4\label{n}]{{\includegraphics[width=0.243\linewidth,height=5cm]{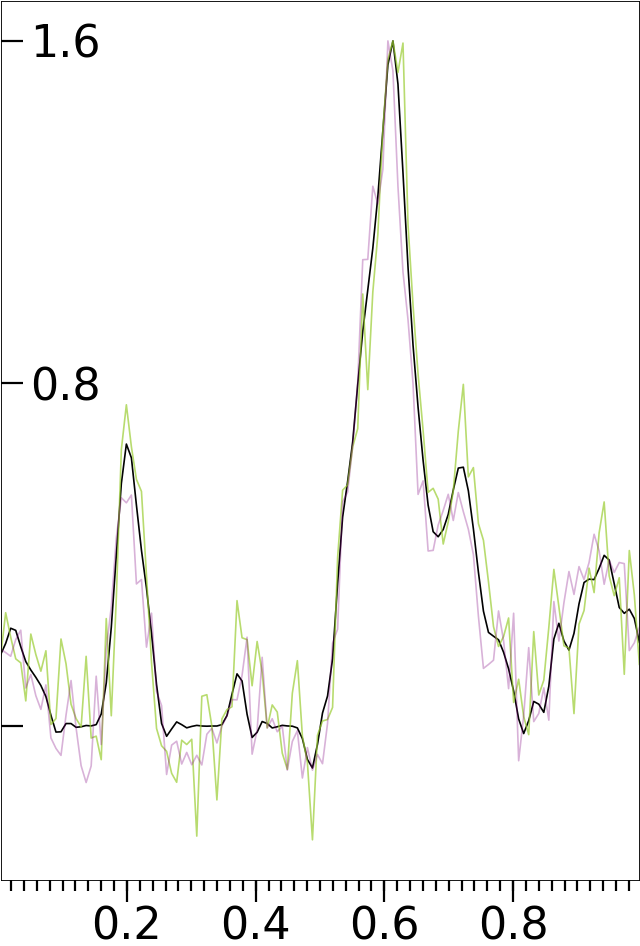} }}
    \caption{Pulse profiles for non-PTA pulsars in band-3 (B3) and band-4 (B4) of the uGMRT. {The profiles in band-3 are normalised by their peak intensity values. In band-4, the normalised profiles are scaled by the ratio of S/N in band-4 to band-3}. The figure style and color schemes are the same as in Figure \ref{PTA_pulsar_profiles}. The central frequency of the {lowest and highest-frequency subbands} of B4 are 575 MHz and 725 MHz, respectively.}
    \label{GMRT_pulsar_profiles}
\end{figure}

We observed 4 GMRT discovered pulsars (from now on we will refer them as ``non-PTA pulsars") {J1120-3618, J1646-2142, J1828+0625, and J2144-5237 (\cite{2019ApJ...881...59B} and \cite{2022ApJ...933..159B})}. These pulsars were observed in band-3 and band-4. PTA pulsars {J1640+2224, J1713+0747, J1909-3744, and J2145-0750 (\cite{2021ApJS..252....4A}, \cite{2021ApJS..252....5A})} were observed in band-3 and band-5.
Table \ref{table_flux} lists the period, DM, and flux densities of these eight MSPs. {Figures \ref{PTA_pulsar_profiles} and \ref{GMRT_pulsar_profiles} show the pulse profiles of PTA and non-PTA pulsars, respectively, for the {lowest} and {the highest-frequency subbands} of the observing bands.}
\begin{table}[H]
\begin{center}
\begin{tabular}{|c|c|c|c c c|}
 \hline
  &&Dispersion&\multicolumn{3}{c|}{Flux density} \\*
  PSR&Period&Measure& 400 MHz & 650 MHz & 1260 MHz \\*
  &(ms)&($\,$pc~cm$^{-3}$)& \multicolumn{3}{c|}{(mJy)}\\*
  \hline\hline
  {J1120$-$3618} & 5.56 & 45.13 & 0.6 & - & -  \\*
  {J1646$-$2142} & 5.85 & 29.74 & 2.2 & 1.1 & -  \\*
  {J1828$+$0625} & 3.63 & 22.42 &1.3 & - & -  \\*
  {J2144$-$5237} & 5.04 & 19.55 & 1.2 & 0.6 & - \\*

  \hline

  J1640$+$2224 & 3.16 & 18.43 & 21.2 &-& 0.8 \\*
  J1713$+$0747 & 4.57 & 15.98 & 6.4 &-& 10.0 \\*
  J1909$-$3744 & 2.95 & 10.39 & 3.8 &-& 0.5 \\*
  J2145$-$0750 & 16.05 & 9.00 & 23.8 &-& 7.5 \\*

  \hline
  \end{tabular}
  \caption{The table lists the basic parameters of the observed pulsars. The flux density values for the non-PTA pulsars {(first four)} are from \cite{2019ApJ...881...59B} and \cite{2022ApJ...933..159B}. {We measured the flux density values of the PTA pulsars (last four) from our observations. We also estimated the flux value of J2144$-$5237 at 650 MHz (not available earlier) from the current data.}}
  
  \label{table_flux}
  \end{center}
  \end{table}

 The four PTA pulsars are some of the best-timed MSPs and were chosen  based on their high detection significance at 322/607 MHz with legacy GMRT as reported by \cite{2020arXiv200908409J}. Along with band-3, PTA pulsars were also observed in band-5 with uGMRT with maximum possible sensitivity to compare the ToA and DM precision with the values reported by \cite{2021ApJS..252....5A} at 1.4 GHz. 
 
 The non-PTA pulsars were selected from the set of GMRT discovered MSPs which have good S/Ns ($>$30 in 40$-$55 minutes for most of them) in band-3. Table \ref{table_1} shows the S/N of the observed MSPs in different frequency bands of the uGMRT. Among the non-PTA pulsars, J1646$-$2142 and J2144$-$5237 are also bright in band-4. J1120$-$3618 and J1828$+$0625 are having relatively lower detection significance in band-4, thus their band-4 observations are excluded from this work.\\*
 \begin{table}[H]
 \begin{tabular}{||c|c|c|c|c||}
 \hline
 MSP & Mean Observation & Median S/N & No. of & Timing base-\\ [0.5ex]
 & time (mins) & Band 3~~~Band 4~~~Band 5 & Epochs &-line (years)\\
 \hline\hline
 J1120$-$3618 & 50 & 70~~~~~~~~~-~~~~~~~~~-- & 13 &3.25\\
 J1646$-$2142 & 40 &  80~~~~~~~~40~~~~~~~~-- & 36 &4.08\\
 J1828$+$0625 & 40 & 30~~~~~~~~~-~~~~~~~~~-- & 15 &2.08\\
 J2144$-$5237 & 55 &  50~~~~~~~~80~~~~~~~~-- & 32 &4.00\\
 \hline
  J1640$+$2224 & 25 &  400~~~~~~~~~-~~~~~~~~~60 & 12 &0.75\\
 J1713$+$0747 & 25 &  160~~~~~~~~~-~~~~~~~~250 & 12 &0.75\\
 J1909$-$3744 & 20 &  120~~~~~~~~~-~~~~~~~~~20 & 7 & 0.75\\
 J2145$-$0750 & 30 &  2600~~~~~~~~-~~~~~~~~200 & 15 &0.75\\
 \hline
\end{tabular}
\caption{The observational properties for the non-PTA pulsars (top four) and PTA pulsars (bottom four) used in this work, along with their detection significance. Fourth column lists the number of observations taken for each pulsar. In this counting, an incoherently dedispersed observation was ignored whenever simultaneous coherently dedispersed data was available in the same band.}
\label{table_1}
\end{table}

\begin{table}[H]
    
    \centering
    \begin{adjustbox}{width=\columnwidth,center}
    \begin{tabular}{|c|c|c|c|c|c|c|c|}
    \hline
    uGMRT &Mode & Frequency& Usable& Time& No. of\\
    Band &    & range (MHz) & bandwidth (MHz) & Resolution ($\mu s$) & Antennas\\
    \hline
    Band-3& I & 300-500 & 135 & 81.92 & 22\\
    Band-4& I & 550-750 & 152 & 81.92 & 25\\
    Band-5& I & {1060}-1460& 300 & 81.92 & 27\\
    \hline
    Band-3 & C & 300-500 & 135 & 10.24/20.48/40.96$^{\dagger}$& 22\\
    Band-4 & C & 550-750 & 152 & 10.24/20.48/40.96$^{\dagger}$ & 25\\
    \hline
    \hline

    \end{tabular}
    \end{adjustbox}
    \caption{The table lists the details of the observational setup in different modes. C and I represent coherent and incoherent dispersion modes. In I mode, filterbank files have 4096 channels in all the bands.\\ $\dagger$ - In C mode, filterbank files have 512/1024/2048 channels in our observations. The table shows the time resolution corresponds to the filterbank with different numbers of channels.}
    \label{Observational_set_up}
\end{table}

Since the aim was to observe with the maximum time-domain sensitivity of the uGMRT, we have taken single sub-array observations, where 70\% and 80\% of the GMRT array was phased and combined to form a single dish with an equivalent gain of 7 K/Jy and 8 K/Jy in band-3 and band-4 respectively. In band-5, we observed with 90$\%$ of the array providing a phased array beam with a gain of 5.9 K/Jy. The observational set up in all frequency band/modes with time resolution, bandwidth, and number of antennas used in phased array are provided in Table \ref{Observational_set_up}. The phased array beam of the uGMRT was recorded after online coherent dedispersion (where each subband voltage samples are corrected for dispersive delays due to the ISM), for band-3 and band-4. In parallel, Stokes-I filterbank data was also acquired for offline incoherent dedispersion (where the intra-channel dispersion smearing is not corrected).  In band-3, we mask the 360$-$380 MHz frequency band affected by the persistent Mobile User Objective System emission. In band-5, Stokes-I filterbank data was acquired for offline incoherent dedispersion. The online coherent dedispersion mode is currently not available for 400 MHz observational bandwidth.

The intra-channel dispersion smearing of the incoherently dedispersed data is decided by 4096 spectral channels over 200 MHz bandwidth in band-3. For example, in case of J1120$-$3618 pulsar having the highest DM in our sample, the intra-channel smearing is 0.28 ms. The uGMRT observations (both coherently dedispersed and raw filterbank) were incoherently dedispersed with known DM value to remove the inter-channel dispersive delays. We performed the incoherent dedispersion and folding of filterbank file using the \texttt{PREPFOLD} command available in \texttt{PRESTO} \citep{2011ascl.soft07017R}. For the purpose of data reduction, we used the ephemeris from legacy GMRT timing studies (\cite{2019ApJ...881...59B} and \cite{2022ApJ...933..159B} for the non-PTA pulsars and NANOGrav ephemeris\footnote{\label{NanoGrav_archives}\url{https://data.nanograv.org/}} {(the latest from 2020/2021 is the NANOGrav 12.5-year data set, version 4; \cite{2021ApJS..252....4A}, \cite{2021ApJS..252....5A})} for the PTA pulsars. We converted \texttt{PREPFOLD} folded data cubes to FITS format for further analysis using \texttt{PAM}  command available in \texttt{PSRCHIVE} \citep{2012AR&T....9..237V}. For timing analysis, we divided band-3 into 128 frequency subbands for the PTA pulsars and 16 subbands for non-PTA pulsars. The band-4 and band-5 were divided into 16 subbands for non-PTA and PTA pulsars.

\begin{figure}[H]
    \centering
    \includegraphics[width=16cm, keepaspectratio]{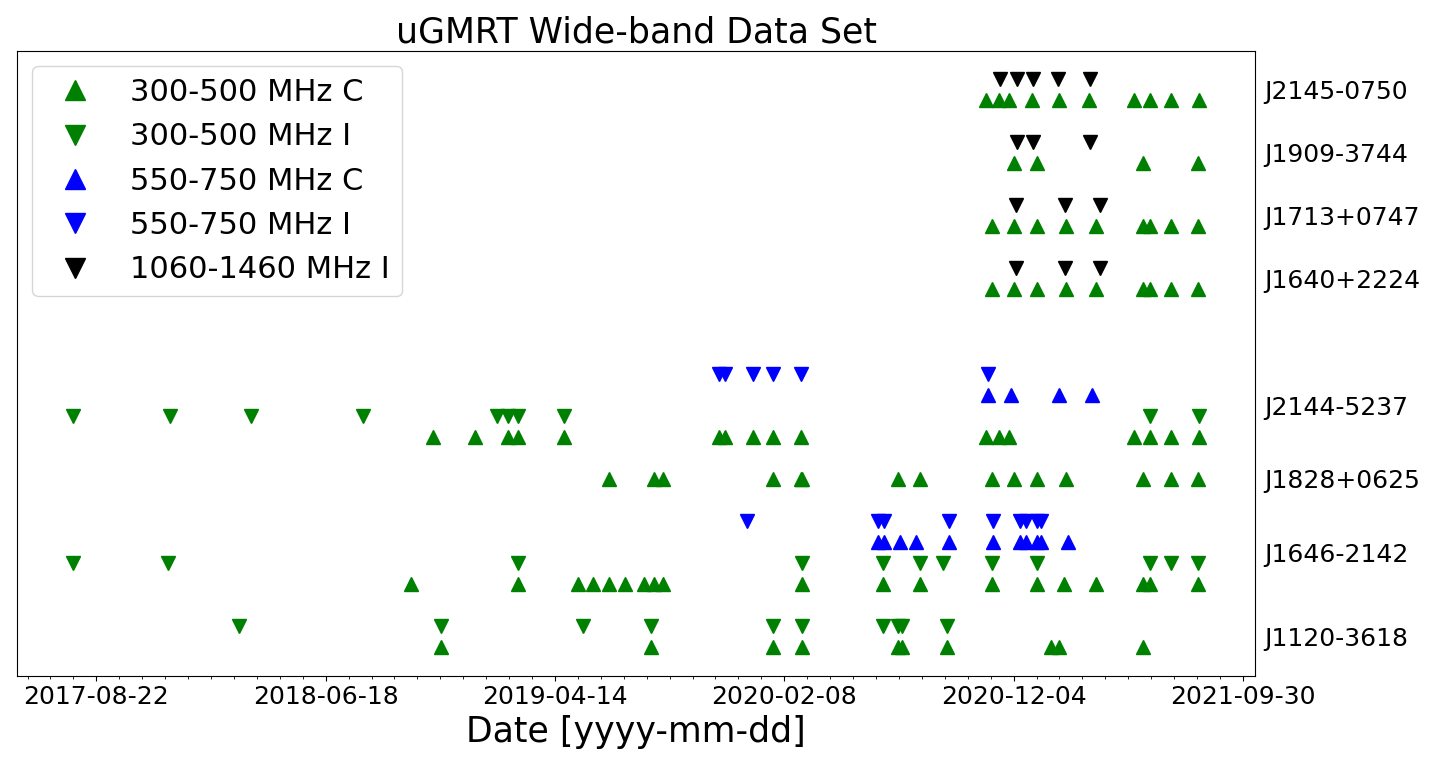}
    \caption{The cadence and length of the observing campaign for each pulsar in different uGMRT bands (marked by separate colours) for different observing modes, where C stands for observations with online coherent dedispersion and I for offline incoherent dedispersion. Note that table \ref{table_1} lists simultaneous C and I observations, presented in different colour points here, as a single epoch observation.}
    \label{fig:1}
\end{figure}

 The four PTA pulsars were observed once a month and covered a span of $\sim$ nine months, while more than 2 years of data were available for all non-PTA pulsars. Fig. \ref{fig:1} shows the cadence for all observed pulsars in the different frequency bands. We also included incoherently dedispersed data available for MSPs J1120$-$3618, J1646$-$2142 and J2144$-$5237 in band-3 and band-4 taken before online coherent dedispersion mode was established.

\section{Timing techniques}
\label{sec:techniques}
\subsection{Narrow-band timing}
Templates generation procedure: In NB timing analysis, all the coherently dedispersed FITS files in a given uGMRT band, with significant pulse detection, are aligned using frequency invariant phase offsets. Similar procedure is applied for incoherently dedispersed FITS files of the same band independently. The aligned FITS are added for a particular frequency band and then averaged in frequency, separately for coherently and incoherently dedispersed FITS, to create a reference template profile for the same band. We used different templates for different frequency bands of the uGMRT, also, independent templates are used for coherently and incoherently dedispersed FITS. Gaussians are fitted, using \texttt{PAAS} command in \texttt{PSRCHIVE} \citep{2012AR&T....9..237V}, to these frequency-time averaged profiles to create analytic noise-free templates. 

NB ToA and DM estimations: For each individual uGMRT frequency band we keep the intra-band frequency resolution in FITS files (16/128 subbands, section \ref{sec:Observations})  and extracted 16/128 ToAs for each epoch FITS in band-3 or band-4 or band-5. Coherently and incoherently dedispersed FITS are dealt separately using the same procedure.

\cite{1992PTRSL.341..117T} prescribes a Fourier frequency-domain technique for measuring the phase shifts between data profile and template by applying cross-correlation between them in the Fourier domain. The estimation of phase-shifts in Fourier domain ensures that the ToA precision is not limited by phase bin resolution. All NB ToAs are estimated using this technique as described in Appendix A of \cite{1992PTRSL.341..117T}. DM is fitted individually for each epoch keeping other parameters fixed in TEMPO2 \citep{2012ascl.soft10015H} to measure the temporal variation of DM. 

\subsection{Wide-band timing}
For WB analysis, we select FITS files having high S/N, with same central frequency to create the profile templates. Different WB templates are used for separate frequency bands and observing modes. We have flagged a few start and end channels plus the channels with bad RFI condition in each band. Here we provide a brief description of ToA and template creation procedure in WB timing, and refer to \cite{2014ApJ...790...93P} and \cite{2019ApJ...871...34P} for further details.

Templates generation procedure: Considering the observations with one dedispersion mode, in one frequency band, the FITS data having highest S/N is used as the initial phase alignment reference. The FITS for each epoch are then aligned relative to that initial alignment reference by determining a constant offset between them and an offset proportional to $\nu^{-2}$ fitted over 16/128 subbands, where the $\nu^{-2}$ factor accounts for DM variability from one observation to another. The aligned FITS of all epochs are averaged together while keeping the 16/128 subbands frequency resolution. Upon iteration, it uses that result as the new reference for alignment and the process is repeated multiple times to create an ``average portrait''. Then the average portrait is decomposed by Principal Component Analysis (PCA) to find a set of basis eigenvectors such that their linear combination (including mean profile) can result in a frequency-dependent profile template.
        Mean profile ($\tilde{p}_{mean}$) and the basis-eigenvectors ($\hat{e}_i$) are smoothed in the process.
        A template $P(\nu)$ at a particular frequency $\nu$ can be created using the equation:
        \begin{equation}
        P(\nu)=\sum_{i=1}^{n_{eig}}\sum_{j=1}^{n_{B}}c_{ij}B_{tk,j}(\nu)\hat{e}_i+\tilde{p}_{mean}\end{equation}
        where, the first sum runs over the number of basis-eigenvectors $n_{eig}$ and the second over $n_B$, the number of basis splines used in the fit. $\sum_{j=1}^{n_{B}}c_{ij}B_{tk,j}(\nu)$ are the coefficients of the eigenvectors which can evolve with frequency to capture the profile evolution. The default constraint in the software, to limit the number of eigenvectors, is to set a threshold value for S/N of eigenvectors. To determine the threshold value, we first allowed all possible eigenvectors with positive S/N. Then we check each eigenvector's shape and profile evolution of coefficient of eigenvector with frequency. For most eigenvectors, which capture noise, have low S/N with their shapes seem to be noise-like, and, more importantly, their coefficient shows a random behavior with frequency. It leads us to decide S/N cut-offs to determine a reduced set of orthogonal basis-eigenvectors. The coefficients of these reduced eigenvectors exhibit smooth variations with frequency. However, low level variations of profile with frequency can get excluded from the WB template with rejection of low S/N eigenvectors. All of the selected eigenvectors in this work have S/N greater than 50. Note that,  alignment error in the data sets can lead to smearing in the average portrait resulting in more number of eigenvectors. To avoid misalignment we use only clean data sets (with high S/N) to create average portrait and select the initial guess as the highest S/N epoch data. For the observed pulsars we find that only 0-2 components were fitting the above criteria for basis-eigenvectors. 
        
        WB ToA and DM estimation: We use the required notation as described in \cite{2014ApJ...790...93P}, \cite{2019ApJ...871...34P} and include them here for completeness. 
Assuming a time-domain model of the pulse profile at frequency $\nu$ has the form 
\begin{equation}
    D(\nu, \phi)=B(\nu)+a(\nu)P(\nu, \phi-\phi(\nu))+N(\nu, \phi)
\end{equation}
(\cite{1992PTRSL.341..117T}; \cite{2014ApJ...790...93P}), where $\phi$ represents the rotational
phase of the pulsar at a given time. $a(\nu)$ and $\phi(\nu)$ are the required scale and shift of the data $D(\nu, \phi)$ from a template $P(\nu, \phi-\phi(\nu))$. $B(\nu)$ is the effective band-pass shape of the receiver and $N(\nu, \phi)$ is additive noise assumed to be normally distributed in the absence of RFI.

The receiver band-width is divided into $n$ frequency channels with the $n$th channel having central frequency $\nu_n$.
A pulse profile at frequency $\nu_n$ is sampled into equally spaced intervals in pulse-phase. The one-dimensional Discrete Fourier Transform (DFT)  of equation (3), after discretising the terms, with respect to rotational phase $\phi$, and making use of the discrete Fourier shift theorem leads to
\begin{equation}
    d_{nk}=a_n p_{nk}e^{-2\pi i k \phi_n} + n_{nk},
\end{equation}
where $k$ refers to the $k$th Fourier frequency of the DFT, $n$ is $n$th frequency channel. $d_{nk}$, $p_{nk}$ and $n_{nk}$ are the DFT of the data $D$, the template $P$ and the noise 
$N$ defined in {equation (4)}. 
Minimising 
        \begin{equation}
             \chi^2(\phi_n, a_n)=\sum_{n,k}\frac{|d_{nk}-a_np_{nk}e^{-2 i \pi k \phi_n}|^2}{\sigma^2_n}
             \label{chi_sq_eqn_WB}
        \end{equation}
        will provide estimates of the scaling factor $a_n$ and phase shift $\phi_n$ between the data $d_{nk}$ and the template $p_{nk}$. Each term in the sum is weighted by the square of the noise $\sigma_n$ estimated in $d_n$. This approach is similar to the technique described in \cite{1992PTRSL.341..117T}, but in this case, the template can evolve with frequency. The WB technique is enabled to simultaneously measure the ToA and DM by inclusion of the constraint in the equation below
        \begin{equation}
            \phi_n(\nu_n)=\phi_o+\frac{K\times DM}{P_s}(\nu_n^{-2}-\nu_{\phi_o}^{-2})
            \label{equation_fit}
        \end{equation} 
        where $\phi_o$ is the phase offset estimated at reference frequency $\nu_{\phi_o}$. $P_s$ is the period of the pulsar, $K$ is the dispersion constant, and $\nu_{\phi_o}$ is a choice of parameterization. The \textit{PulsePortraiture} gives freedom to select the value of $\nu_{\phi_o}$. However, we have used the default feature of the package which estimates the value of $\nu_{\phi_o}$ for zero covariance between $\phi_{o}$ (WB ToA) and DM. WB results in a single ToA and DM for each epoch. We have analysed individual frequency bands and modes (coherently and incoherently dedispersed) FITS files separately for ToA and DM estimations using different templates.
        
                A sample of WB timing analysis ``jupyter notebooks'' developed for the uGMRT band-3, band-4, and band-5, are available in the github\footnote{\url{https://github.com/Shyamss6027557/Wide-band-timing-at-Low-frequencies-with-uGMRT}}. It needs a folded data cube in PSRFITS format. The WB ToAs and DM values for PTA MSPs from these GMRT observations are also provided there.

        Over the uGMRT frequency bands, the pulse profiles show a clear difference in profile shapes from one band to another. So the absolute DM values are expected to be different for the two non-simultaneous bands. Also, the $\nu^{-2}$ fitting for DM estimation in NB analysis captures part of profile evolution with the frequency that could result in different absolute DM values in the NB and WB analysis. To account for the DM variability, we've subtracted the weighted mean of DM values from estimated DM of individual epochs, separately for a particular band, observing mode, and analysis (NB and WB).
\section{Results}
\label{sec:Results}
With the aim to validate the WB timing pipeline we have carried out a comparative study of NB and WB timing for the PTA pulsars {and then the validated pipeline is applied to the non-PTA pulsars.} 
\subsection{Validation of the WB timing pipeline by PTA pulsars} 
We have created 1024 bins for the band-3 profile of J2145$-$0750, and the rest of the PTA pulsars have 256 profile bins in band-3. In Band-5, we have created profiles with 64 bins resolution. For the PTA pulsars, we have not fitted for long-term timing model (except DM for individual epochs), due to the availability of shorter timing span ($<$ 1 year) with sparse sampling. The ephemeris, for the PTA pulsars, are obtained from the NANOGrav archive$^{\ref{NanoGrav_archives}}$.
\subsubsection{DM variation for PTA pulsars}

\begin{figure}[H]
\centering
        \includegraphics[width=\linewidth,keepaspectratio]{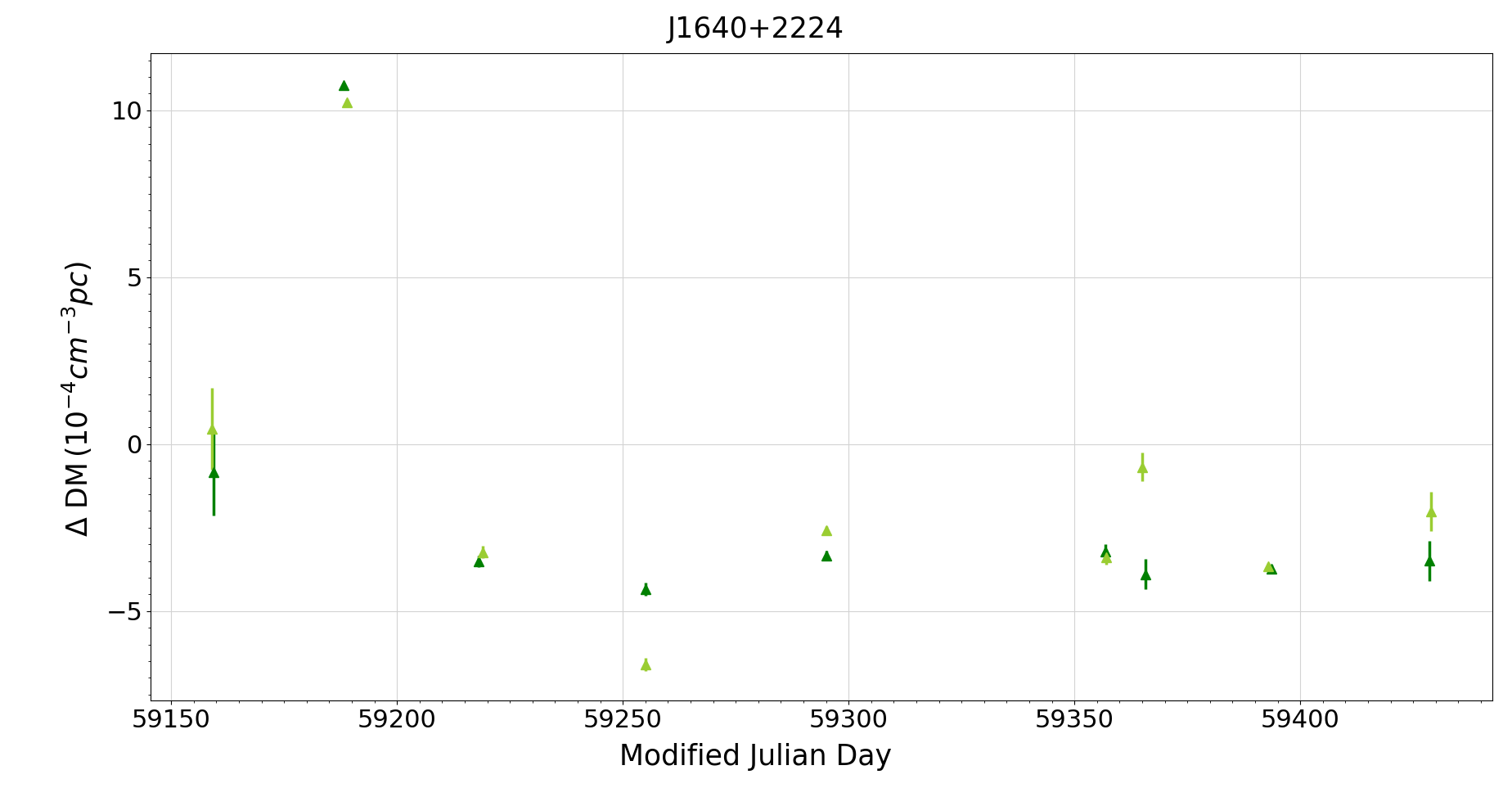}
        \vspace{-10mm}
    \caption{{Figure showing DM variation with time for the PTA pulsar J1640$+$2224 in band-3 of the uGMRT. Green and yellow-green triangles show the DM values obtained in WB and NB analysis, respectively.}}
    \label{DM_curve_J1640+2224}
\end{figure}

\begin{figure}[H]
\centering
        \includegraphics[width=\linewidth,keepaspectratio]{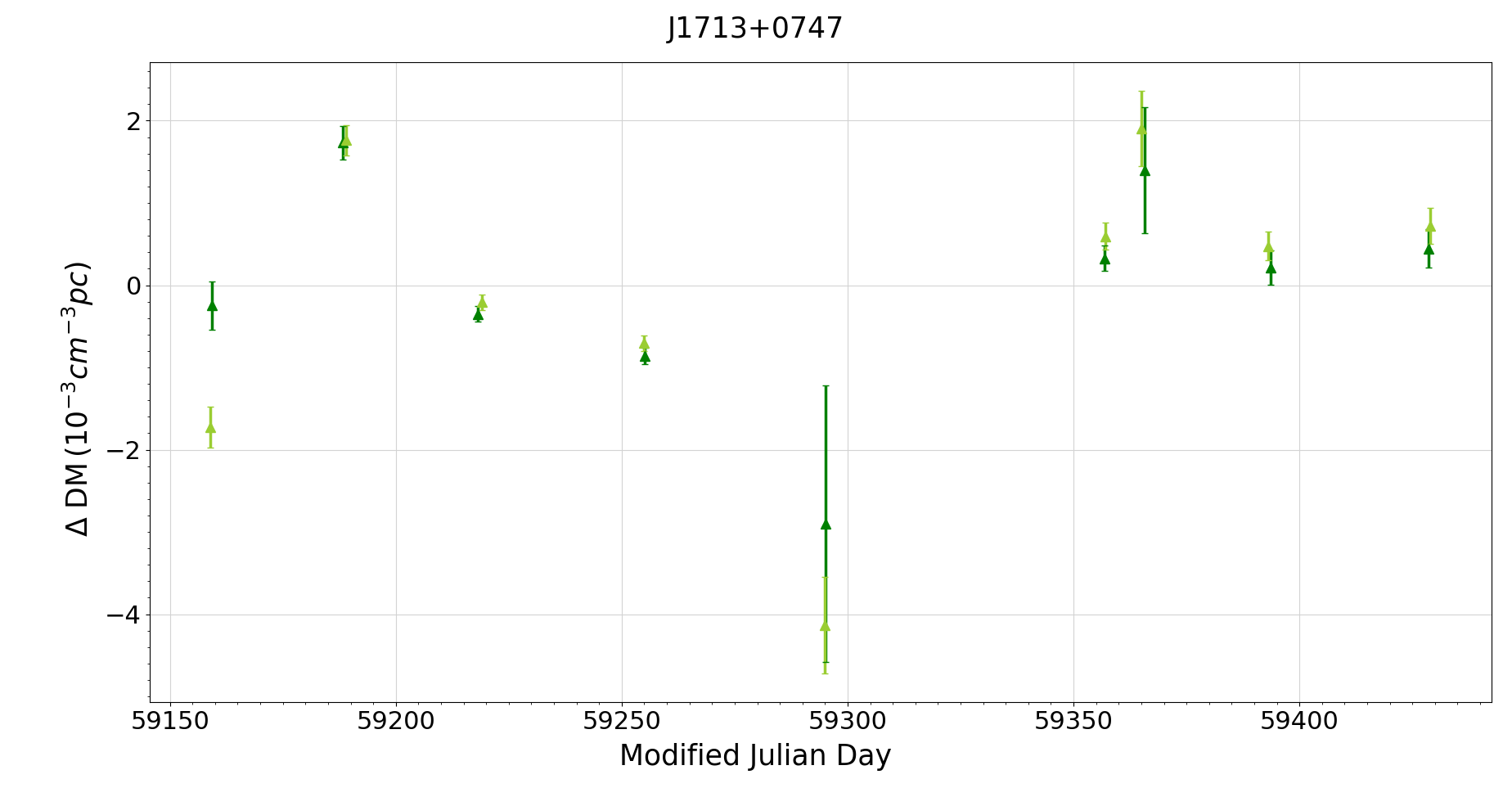}
        \vspace{-10mm}
    \caption{{Figure showing DM variation with time for the PTA pulsar J1713$+$0747 in band-3 of the uGMRT. The colour schemes are same as in Figure \ref{DM_curve_J1640+2224}.}}
    \label{DM_curve_J1713+0747}
\end{figure}

\subsubsubsection{J1640$+$2224}
{Figures \ref{b} and \ref{f}} show folded pulse profiles for J1640$+$2224 in band-3 and band-5, respectively. The steep spectral nature of this pulsar makes it much brighter in band-3 as compared to band-5. {Figure \ref{DM_curve_J1640+2224} shows the DM variation with time for this pulsar in band-3}. 
\subsubsubsection{J1713$+$0747}
J1713$+$0747 has a lower flux density in band-3 compared to band-5 (Table \ref{table_flux}). The pulse profiles in both the bands have a single component with significant pulse broadening due to scattering seen in band-3 (Figure \ref{d}) compared to band-5 (Figure \ref{h}). {The pulse profile within band-3 also evolves considerably with frequency. For J1713$+$0747, we measured a scintillation bandwidth of 0.85$\pm$0.22 MHz at 334 MHz and 1.24$\pm$0.23 MHz at 425 MHz. We find $\Delta \nu \propto \nu^{1.56}$ where $\Delta \nu$ is scintillation bandwidth at frequency $\nu$. The estimated scaling is much shallower than the Kolmogorov spectrum. The coefficients of eigenvectors, created in PCA analysis, capture the profile evolution (including scattering) with frequency.} {Figure \ref{DM_curve_J1713+0747} shows the DM variation with time for this pulsar in band-3.} 

\subsubsubsection{J1909$-$3744}
 J1909$-$3744 has a single component pulse profile both in band-3 (Fig \ref{c}) and band-5 (Fig \ref{g}), and has higher detection significance in band-3 compared to band-5. {Due to less number of observations (only four) in band-3 for J1909$-$3744 the temporal variation of DM plot is not added. However, the median precision obtained for the available epochs are listed in Table \ref{Table_2}}.

\begin{figure}[H]
\centering
        \includegraphics[width=\linewidth,keepaspectratio]{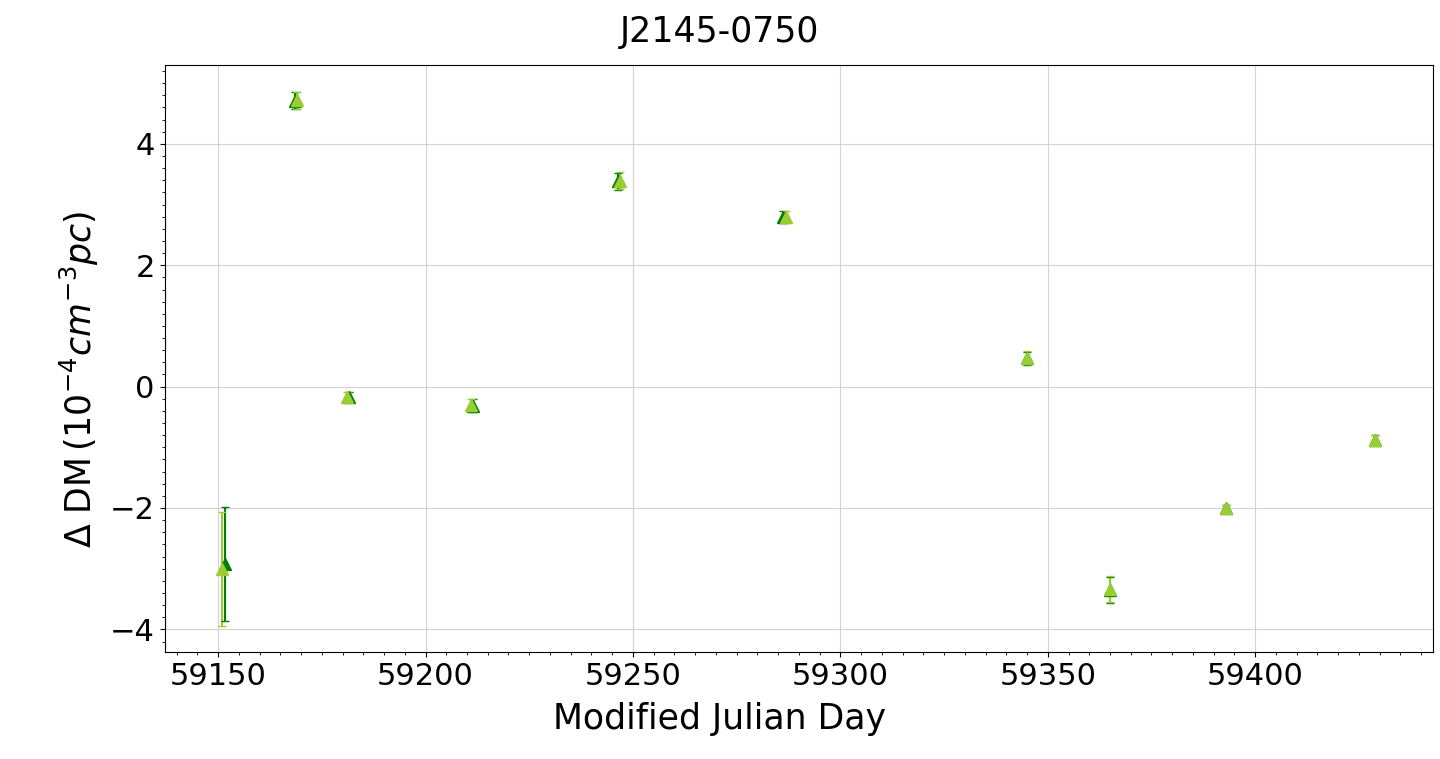}
        \vspace{-10mm}
    \caption{{Figure showing DM variation with time for the PTA pulsar J2145$-$0750 in band-3 of the uGMRT. The colour schemes are same as in Figure \ref{DM_curve_J1640+2224}.}}
    \label{DM_curve_J2145-0750}
\end{figure}

\subsubsubsection{J2145$-$0750}
The pulse profile of J2145$-$0750 has two main components, and the peak amplitude ratio evolves  with frequency (as seen in Fig. \ref{a} and \ref{e}). The pulsar is bright in band-3, making it one of the best PTA MSPs to follow-up at the low frequencies. Figure \ref{DM_curve_J2145-0750} presents the temporal variations of DM for J2145$-$0750, obtained with NB and WB analysis, {in band-3 of the uGMRT}.

\subsubsection{Comparison between NB and WB timing for PTA pulsars}

{The PTA pulsars show similar temporal DM variations for NB and WB analysis for most of the epochs. Significant temporal variation of DM ($>$ $\pm$3$\sigma_{DM}$) is seen for the PTA pulsars.}

\begin{figure}[H]
    \centering
   \includegraphics[width=\linewidth, keepaspectratio]{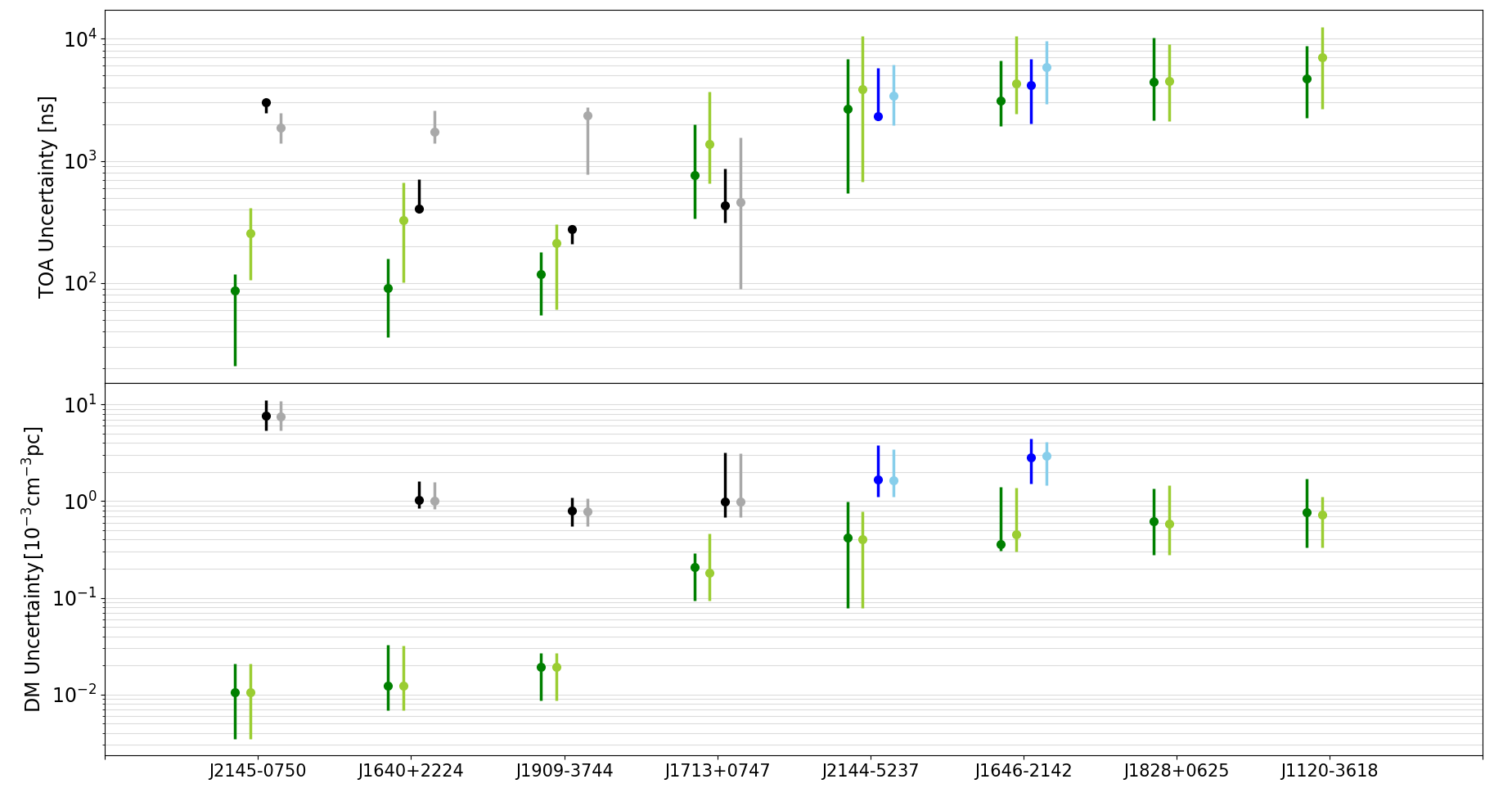}
    \caption{Figure showing median ToA (top panel) and DM (bottom panel) uncertainties obtained in the two (NB and WB) analyses for eight pulsars. Error bars represent the range of precision obtained for the individual pulsar data sets. {Pulsars are arranged on the x-axis in increasing order of their band-3 DM uncertainty.} We used {green, blue, and black colours} to represent the values obtained from WB analysis in band-3, 4, and 5, respectively. Similarly, {light-green, sky-blue, and dark-gray} colours are used for NB analysis values in band-3, 4, and 5, respectively.}
    \label{toa_dm_NB}
\end{figure}

{Table \ref{Table_2} presents a comparison of our results from NB and WB timing analysis which lists raw (not scaled by observing bandwidth and duration) ToA and DM uncertainties. {Figure \ref{toa_dm_NB} shows median ToA and DM precision along with the range of uncertainties (plotted as error bars) in NB and WB analysis, respectively.\iffalse, for all the pulsars in band-3, band-4, and band-5.\fi} In general, ToAs are more precisely estimated in WB than NB timing. The median improvement in ToA uncertainty from NB to WB analysis are 2.4 and 2.7 times in band-3 and band-5, respectively. The ToAs in WB analysis are measured at the zero-covariance frequency ($\nu_{\phi_o}$). The ToA uncertainty has its minimum value at the estimated frequency since at other frequencies there will be some covariance between DM and ToA which will lead to a higher ToA uncertainty. For PTA pulsars, we get median ToA uncertainty $\sim 100$ ns in WB analysis of band-3 observations except J1713$+$0747 (having $\sigma_{ToA}\sim 0.8\,\mu s$ ). In band-5, the median ToA uncertainty is $< 500$ ns in WB analysis for all PTA pulsars except J2145$-$0750 (having $\sigma_{ToA}$ $\sim$ 3 $\mu s$). Band-3 ToAs are at least 2 times more precise than band-5 ToAs except for J1713$+$0747 as its detection significance is higher in band-5 as compared to band-3.

DM precision from WB and NB analysis are almost same for all of the observations both in band-3 and band-5. In band-3, we find the median DM precision of $1-2\times10^{-5}$ $\,pc~cm^{-3}$ for all the PTA pulsars except J1713$+$0747 (having $\sigma_{DM}\sim2\times10^{-4}$ $\,pc~cm^{-3}$). In band-5, the median DM precision ranges from 10$^{-4}$ to 10$^{-3}$ $\,pc~cm^{-3}$. We find a minimum of 5 times improvement in DM precision from band-5 to band-3. In case of NB analysis, we have used the ``norescale" option while fit for DM using TEMPO2. It disables the scaling of output raw DM uncertainties by minimized chi-square. We used this feature to compare the NB uncertainties directly with raw WB uncertainties.

We see a gradual improvement in ToA and DM precision with increase of number of profile bins for PTA pulsars in both NB and WB analysis. For example, the ToA and DM uncertainties for J2145$-$0750 improve by a factor of $\sim$3 in band-3 by increasing the number of bins from 128 to 1024 in WB analysis.

Table \ref{median_dm_table} contains the median DM values obtained from NB and WB analysis. Also, it shows the number of eigenvectors used for each observed pulsar to model its profile. In case of zero eigenvector, WB analysis will be the same as NB analysis. However, the WB ToA is calculated at zero covariance frequency. Also, the NB templates are created outside of \texttt{PulsePortraiture} by Gaussian fitting, which makes the templates different in the case of NB and WB analysis. We used 1-2 eigenvectors to model WB template in band-3. In band-5, we have not used any eigenvector for all PTA pulsars except J1713$+$0747 (requiring one eigenvector).
 The difference in median DM values from NB and WB analysis lies within $\pm1$ $\sigma_{DM}$ for all PTA pulsars except J1640$+$2224 (having DM difference of $4\times 10^{-4}$ $pc\,cm^{-3}$ in band-3 and $3\times 10^{-3}$ $pc\,cm^{-3}$ in band-5).}

\begin{table}[H]
\centering
\begin{tabular}{|c |c |c |c |c |c |c|}
 \hline
 {{PSR}} & uGMRT&{$\sigma_{ToA}$} & {$\sigma_{ToA}$}& {$\sigma_{DM}$} &  {$\sigma_{DM}$} \\ 
 &Band&{{NB}}&{{WB}} & {{NB}}&{{WB}} \\ 
  &&{$\mu s$}&{$\mu s$}& {$10^{-4}\,pc~cm^{-3}$} & {$10^{-4}\,pc~cm^{-3}$}\\ 
  \hline\hline
  {{J1120-3618}} &3& {7.05}  & {4.71}& {7.3} & {7.7} \\
  {{J1646-2142}} & 3& {4.29} & {3.13}  & {4.5} & {3.6} \\
  {{}} & 4& {5.87} & {4.19} & {29.4} & {28.2} \\
  {{J1828+0625}} &3& {4.47}  & {4.03}  & {5.8} & {6.2} \\
  {{J2144-5237}} &3& {3.84} & {2.69}  & {4.0} & {4.2} \\
  {{}} &4& {3.40} & {2.33} &  {16.4} & {16.6} \\

  \hline
  
  {{J1640+2224}} &3& {0.326}  & {0.091}  & {0.12} & {0.12} \\
  {{}} &5& {1.736}  & {0.409} & {10.16} & {10.25} \\
  {{J1713+0747}} &3& {1.366}  & {0.762}  & {1.82} & {2.08} \\
  {{}} &5& {0.461}  & {0.430}  & {9.89} & {9.90} \\
  {{J1909-3744}} &3& {0.213}  & {0.119}  & {0.19} & {0.19} \\ 
  {{}} &5& {2.370}  & {0.278}  & {7.85} & {7.92} \\
  {{J2145-0750}} &3& {0.258}  & { 0.087}  & {0.10} & {0.10} \\
  {{}} &5& {1.873}  & {2.997}  & {74.75} & {76.02}\\
 \hline
 \end{tabular}
 \caption{The table lists the median ToA and DM precision obtained from NB and WB analysis. For NB analysis, single ToA over full band is considered here for the comparison with WB analysis.}
 \label{Table_2}
 \end{table}
\vspace{-5mm}
\begin{table}[H]
    \centering
    \begin{tabular}{|c||c|c||c|c||c|c|}
    \hline
    PSR & \multicolumn{2}{c||}{No. of eigenvectors(WB)} & \multicolumn{4}{c|}{Median DM ($\,pc~cm^{-3}$)}\\*
    
     & \multicolumn{2}{c||}{} & \multicolumn{2}{c}{Band-3} & \multicolumn{2}{c|}{Band-4}  \\*
     \hline
     & Band-3 & Band-4 & NB & WB & NB & WB \\*
     \hline
     J1120-3618 & 1&-&              45.1289(7)&45.1289(8)&-&-\\*
     J1646-2142&1&1&                29.7568(4)&29.7568(4)&29.727(3)&29.729(3)\\*
     J1828+0625 & 0  & - &          22.4162(6) & 22.4165(6) & -&-\\*
     J2144-5237 & 1 & 0 &          19.5502(4)& 19.5501(4)&19.553(2)&19.551(2)\\*
     \hline
     & Band-3 & Band-5 & \multicolumn{2}{c||}{Band-3} & \multicolumn{2}{c|}{Band-5}  \\*
     \hline
     J1640+2224&1&0&18.42803(1)&18.42763(1)&18.429(1)&18.426(1)\\*
     J1713+0747&2&1&15.9770(2)&15.9770(2)&15.981(1)&15.981(1)\\*
     J1909-3744&1&0&10.39051(2)&10.39051(2)&10.3991(8)&10.3922(8)\\*
     J2145-0750&1&0&9.00194(1)&9.00194(1)&9.000(7)&8.996(8)\\*
     
    \hline
    \hline
    \end{tabular}
    \caption{{The table lists the number of eigenvectors used to model the WB templates. Also, it presents the median DM values obtained in NB and WB analysis for each pulsar in the subsequent columns.}}
    \label{median_dm_table}
\end{table}

\subsubsection{Comparison with other high-precision timing studies}
{We have scaled the ToA uncertainties by a factor of $\sqrt(\frac{bandwidth}{100 MHz}\times\frac{duration}{30 min})$ (following the similar scaling as \cite{2021arXiv211206908N} and \cite{2021ApJS..252....5A}) to compare the ToA precision obtained in our work with the values from other high precision timing studies. We have also scaled the raw ToA uncertainties reported in \cite{2021arXiv210105334K}. Table \ref{PTA_pulsars_comparison_for_different_studies} shows a comparison of scaled ToA and raw DM precision obtained from \cite{2021arXiv210105334K} (InPTA), \cite{2021arXiv211206908N} (InPTA), \cite{2021ApJS..252....5A} (NANOGrav), and this work.}

  {\cite{2021arXiv210105334K} reported ToA and DM precision for five PTA pulsars, including J1713$+$0747, J1909$-$3744, and J2145$-$0750, from simultaneous 400$-$500 MHz (with 5 antennas of the uGMRT array) and 1360$-$1460 MHz (with 12 antennas of the uGMRT array) observations using multi sub-array mode of the uGMRT. Recently, \cite{2021arXiv211206908N} reported  DM and ToA precision from WB analysis for the same three pulsars using 10 antennas of the uGMRT array at 300-500 MHz.}

\begin{table}[H]
    \centering
    \begin{adjustbox}{width=\columnwidth,center}
    \begin{tabular}{|c||c|c||c|c|c|}
    \hline
    &\multicolumn{2}{c||}{scaled}&\multicolumn{3}{c|}{scaled}\\*
    PSR&\multicolumn{2}{c||}{$\sigma_{ToA}$ ($\mu s$)}&\multicolumn{3}{c|}{$\sigma_{ToA}$ ($\mu s$)}\\*
    \hline
    &N21&S22&K21&S22&NANOGrav\\*
    \hline
    &\multicolumn{2}{c||}{{300-500}}&{1360-1460}&{1060-1460}&{1147-1765/1151-1885}\\*
    &\multicolumn{2}{c||}{{MHz (GMRT)}}&{MHz (GMRT)}&{MHz (GMRT)}&{MHz (AO/GBT)}\\*
    \hline
    &G$\sim$3.2 K/Jy&G$\sim$7.0 K/Jy&G$\sim$2.6 K/Jy&G$\sim$5.9 K/Jy&G$\sim$9-11 K/Jy (AO)\\*
    &&&&&G$\sim$2.0 K/Jy (GBT)\\*
    
    \hline
    J1640+2224 &   -  & 0.12&  -  &0.75&0.26\\*
    \hline
    J1713+0747 & 0.81& 0.98&1.04&0.79&0.04\\*
    \hline
    J1909-3744 & 0.46& 0.14&1.92&0.45&0.09\\*
    \hline
    J2145-0750 & 1.22& 0.12&2.83& 5.99&0.48\\*
    \hline
    
    &\multicolumn{2}{c||}{$\sigma_{DM}$ $(10^{-4}$ $\,pc~cm^{-3}$)}&\multicolumn{3}{c|}{$\sigma_{DM}$  ($10^{-4}$ $\,pc~cm^{-3}$)}\\*
    \hline

    J1640+2224 &   -  &0.1&     -            &10.2&4.0\\*
    \hline
    J1713+0747 & 0.9 &2.1& 1.0$\times10^{2}$ [2$^\dagger$]&9.9&0.2\\*
    \hline
    J1909-3744 & 0.2 &0.2 &11.0$\times10^{2}$ [5$^\dagger$]&7.9&0.9\\*
    \hline
    J2145-0750& 0.3 &0.1 & 8.0$\times10^{2}$ [3$^\dagger$]  &76.0&6.0\\*
    \hline
    \end{tabular}
        \end{adjustbox}
    \caption{{Table lists the scaled ToA (upper part) and raw DM (lower part) precision for the PTA pulsars from \cite{2021arXiv210105334K} (K21), \cite{2021arXiv211206908N} (N21), \cite{2021ApJS..252....5A} (NANOGrav), and this work (S22). Third and fourth rows contain the frequency range, telescope and gain\protect\footnotemark information.\\
    $\dagger$: Inter-band DM precision reported in K21.}}
    \label{PTA_pulsars_comparison_for_different_studies}
\end{table}
\footnotetext{uGMRT, AO, and GBT gain information are available at\\* \url{http://gmrt.ncra.tifr.res.in/~astrosupp/obs_setup/sensitivity.html},\\* \url{http://www.naic.edu/~astro/RXstatus/Lwide/Lwide.shtml\#gain}, and\\* \url{https://science.nrao.edu/facilities/gbt/proposing/GBTpg.pdf}, respectively.}
\vspace{-5mm}
{The use of 22 antennas of the GMRT in the current band-3 observations allowed us to achieve better ToA and DM precision for J2145$-$0750 than \cite{2021arXiv211206908N}. For J1909$-$3744, the DM precision achieved from both the observing set-ups are similar while the ToAs are more precise in our work. For J1713$+$0747, we notice that the DM uncertainty obtained from the current work is 2-times less precise than the \cite{2021arXiv211206908N}. {It could be attributed to the loss of gain during the current observations due to temporal dephasing in longer baseline antennas. Due to such possible loss of sensitivity, the scaled ToA precision for this pulsar was not improved.} The best median ToA and DM precision obtained from our study in band-3 are around 100 ns and 1$\times$10$^{-5}$ pc cm$^{-3}$ which are $\sim$2 and $\sim$4 times better, respectively, than the earlier
GMRT results.}

{For all the 4 PTA pulsars presented in this work, \cite{2021ApJS..252....5A} reported ToA and DM uncertainties at 1.4 GHz from WB analysis using data from Arecibo (AO) and Green Bank Telescope (GBT) \footnote{\url{http://nanograv.org/telescopes/}}. We compare our band-5 results with \cite{2021arXiv210105334K} and  \cite{2021ApJS..252....5A}.} {With the use of 27 antennas at band-5 of the uGMRT, we could achieve sub-$\mu s$ ToA uncertainties for most of the PTA pulsars (except J2145$-$0750). These values are 3$-$20 times less precised than the NANOGrav measurements. However, with the use of lower bandwidth and less number of antennas the achieved ToA precision in \cite{2021arXiv210105334K} is greater than $1\, \mu s$ for the three common PTA pulsars.} {The full band-5 coverage allowed us to achieve DM precision at the level of 10$^{-3}$ $pc\,cm^{-3}$ or better, which is at least an order of magnitude more precise than the 100 MHz bandwidth observations reported in \cite{2021arXiv210105334K}. Moreover, the low-frequency intra-band DM measurements reported in this paper is significantly more precise than the inter-band DM estimates reported in \cite{2021arXiv210105334K}. The DM precision from NANOGrav observations are 3-50 times better than our measurements.}
{The DM and ToA precision achieved in band-3 of uGMRT are better (or atleast on par) than the L-band observations of NANOGrav for most of the PTA pulsar except J1713$+$0747. 
}

 \vspace{-3mm}
\subsection{Results for non-PTA pulsars}
{For all the non-PTA pulsars, coherently and incoherently dedispersed profiles are created with 128 and 64 bins respectively, both in band-3 and band-4. 
Their ephemeris are obtained from \cite{2019ApJ...881...59B} and \cite{2022ApJ...933..159B}. We have fitted model parameters using TEMPO2 timing software. For all the non-PTA pulsars we achieve phase coherent timing over a baseline of 2-4 years. {Table \ref{timing_solutions} shows the fitted model parameters and timing precision achieved for the non-PTA pulsars.} For timing fit, we have regenerated NB ToAs using frequency and time-averaged profiles resulting in a single ToA per epoch. {We have not fitted the global DM (while fitting other parameters) using the parameter file in TEMPO2. However, for a pulsar showing a larger than $\pm 3\sigma$ DM variation, we corrected each epoch NB ToA by its DM value by adding the $-$\texttt{dm} flag in the timing file. For a smaller than $\pm 3\sigma$ DM variation, ToAs are corrected for a fixed DM value available in the parameter file.} {For all the non-PTA pulsars, band-3 coherently dedispersed data set gives the best ToA precision, so we have used only this set of ToAs for the timing model fit.}
}
\subsubsection{DM variation and post-fit timing residuals for non-PTA pulsars}

\begin{figure}[H]
\centering
        \includegraphics[width=0.9\linewidth,keepaspectratio]{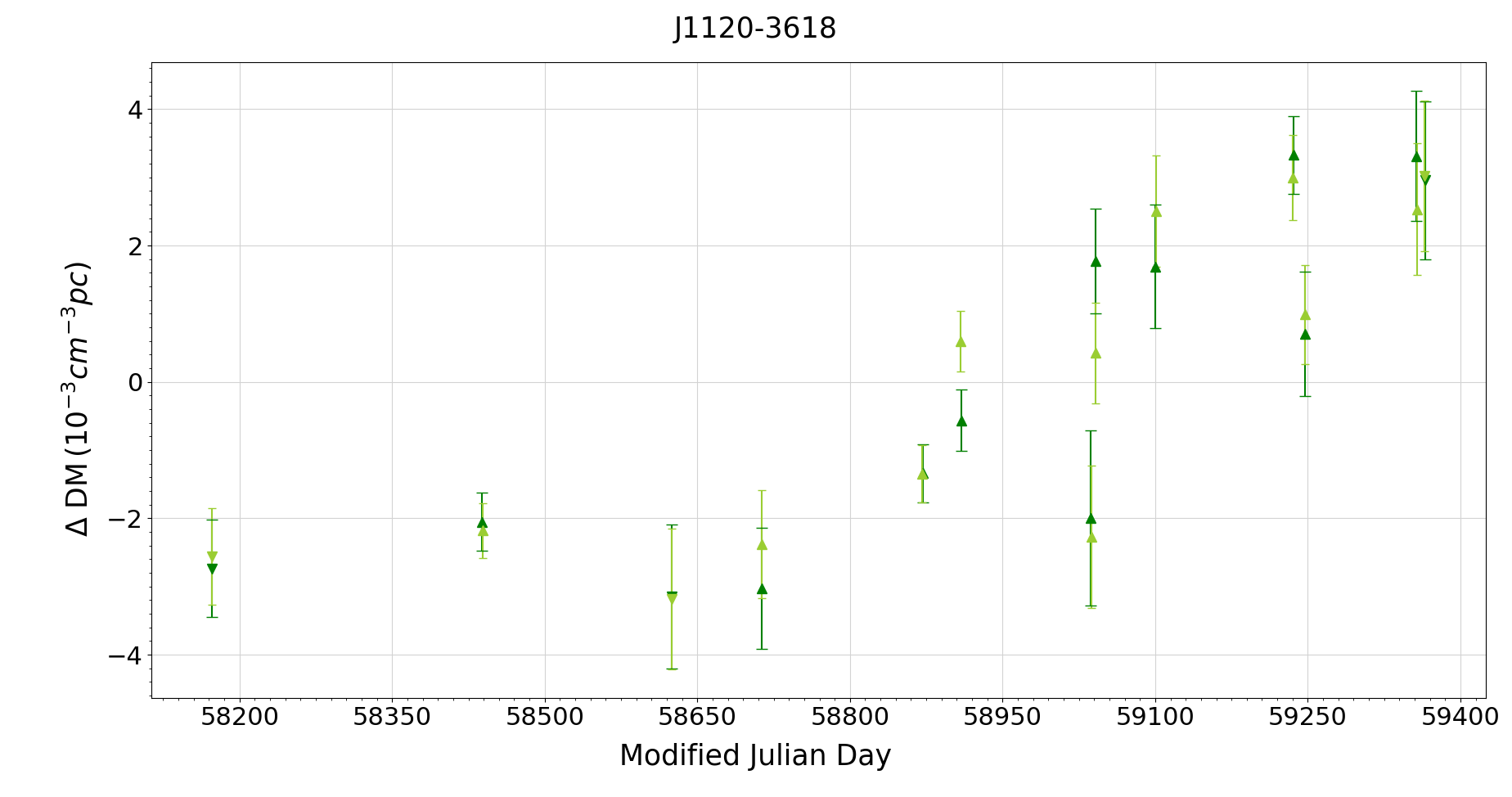}
        \vspace{-5mm}
    \caption{DM variation with time for the non-PTA pulsar J1120$-$3618 in band-3 of uGMRT. {Green} and {Yellow-green} triangles show DM values obtained by WB and NB analysis, respectively. Up and down triangles represent coherently and incoherently dedispersed data.}
    \label{J1120_GMRT}
\end{figure}

\begin{figure}[H]
\centering
        \includegraphics[width=0.9\linewidth,keepaspectratio]{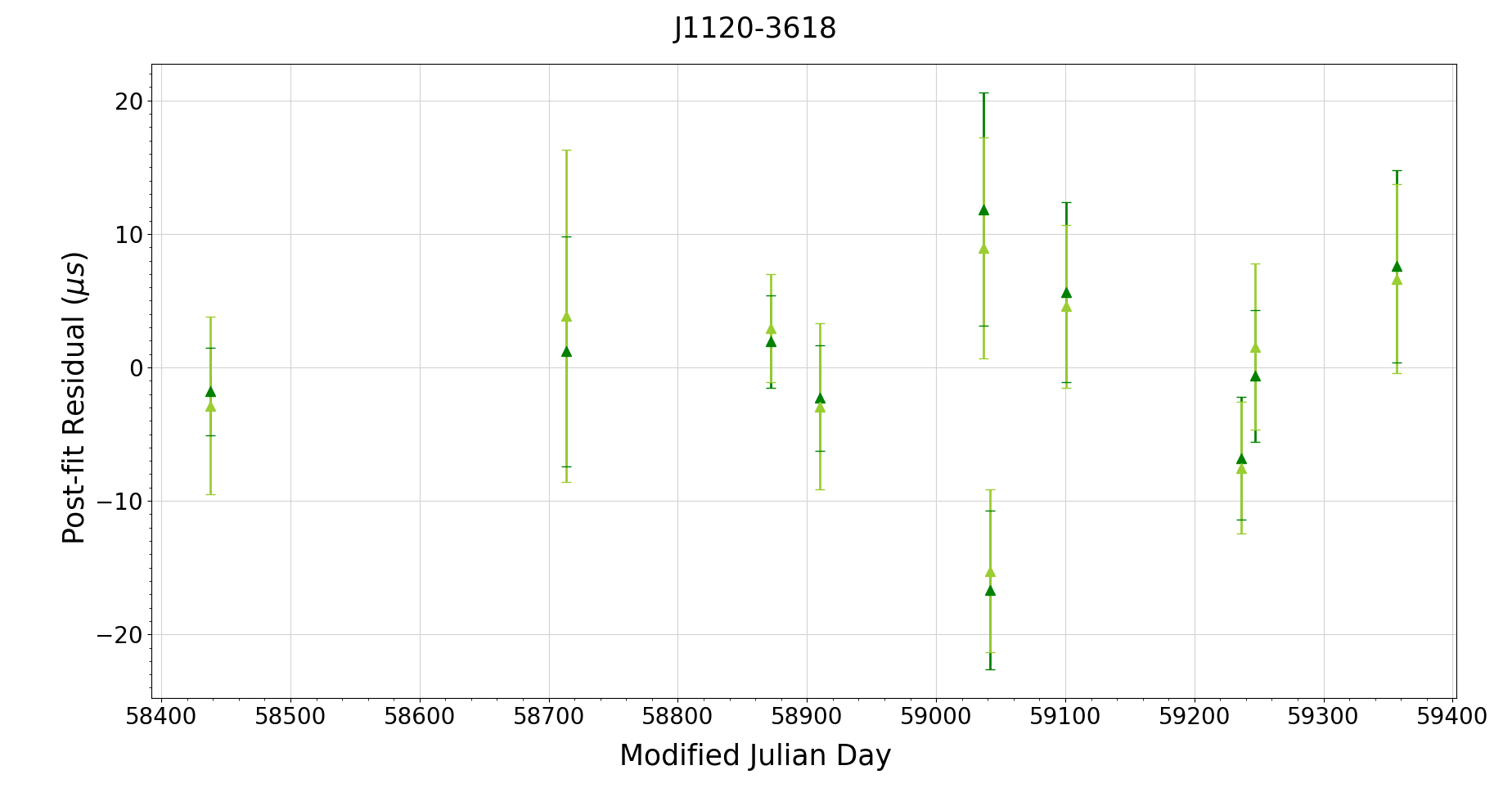}
        \vspace{-5mm}
    \caption{{Post-fit timing residuals versus MJD for the non-PTA pulsar J1120$-$3618. The colour and marker scheme are the same as Figure \ref{J1120_GMRT}.}}
    \label{J1120_GMRT_residuals}
\end{figure}

\begin{figure}[H]
\centering
        \includegraphics[width=0.9\linewidth,keepaspectratio]{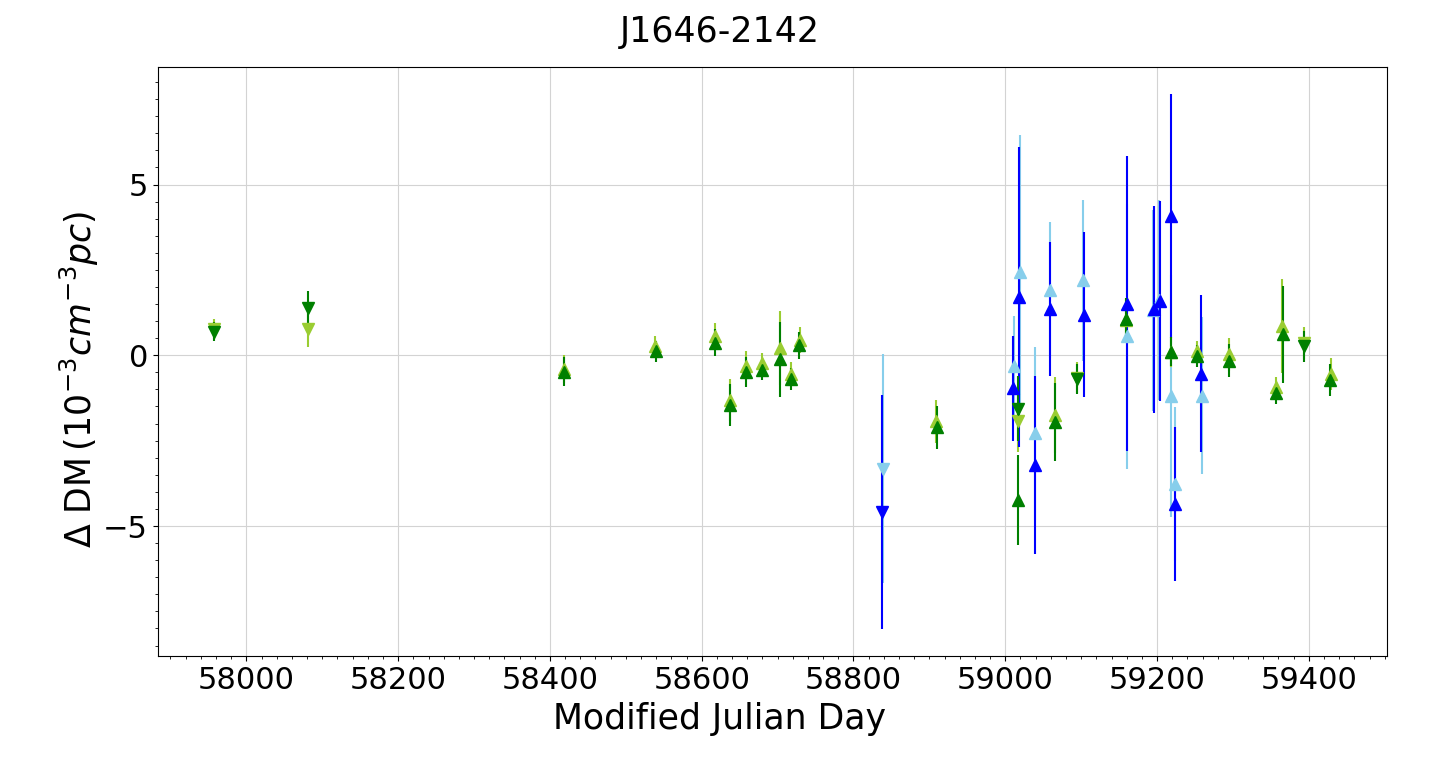}
                \vspace{-5mm}
    \caption{Figure showing DM variation with time for the {non-PTA} pulsar J1646$-$2142 in two separate bands of the uGMRT. {Green and blue} triangles show DM values obtained by WB analysis in band-3 and band-4, respectively. {Yellow-green and sky-blue} triangles show DM values obtained by NB analysis in band-3 and band-4, respectively. Up and down triangles represent coherently and incoherently dedispersed data.}
    \label{J1646_GMRT}
\end{figure}

\begin{figure}[H]
\centering
        \includegraphics[width=0.9\linewidth,keepaspectratio]{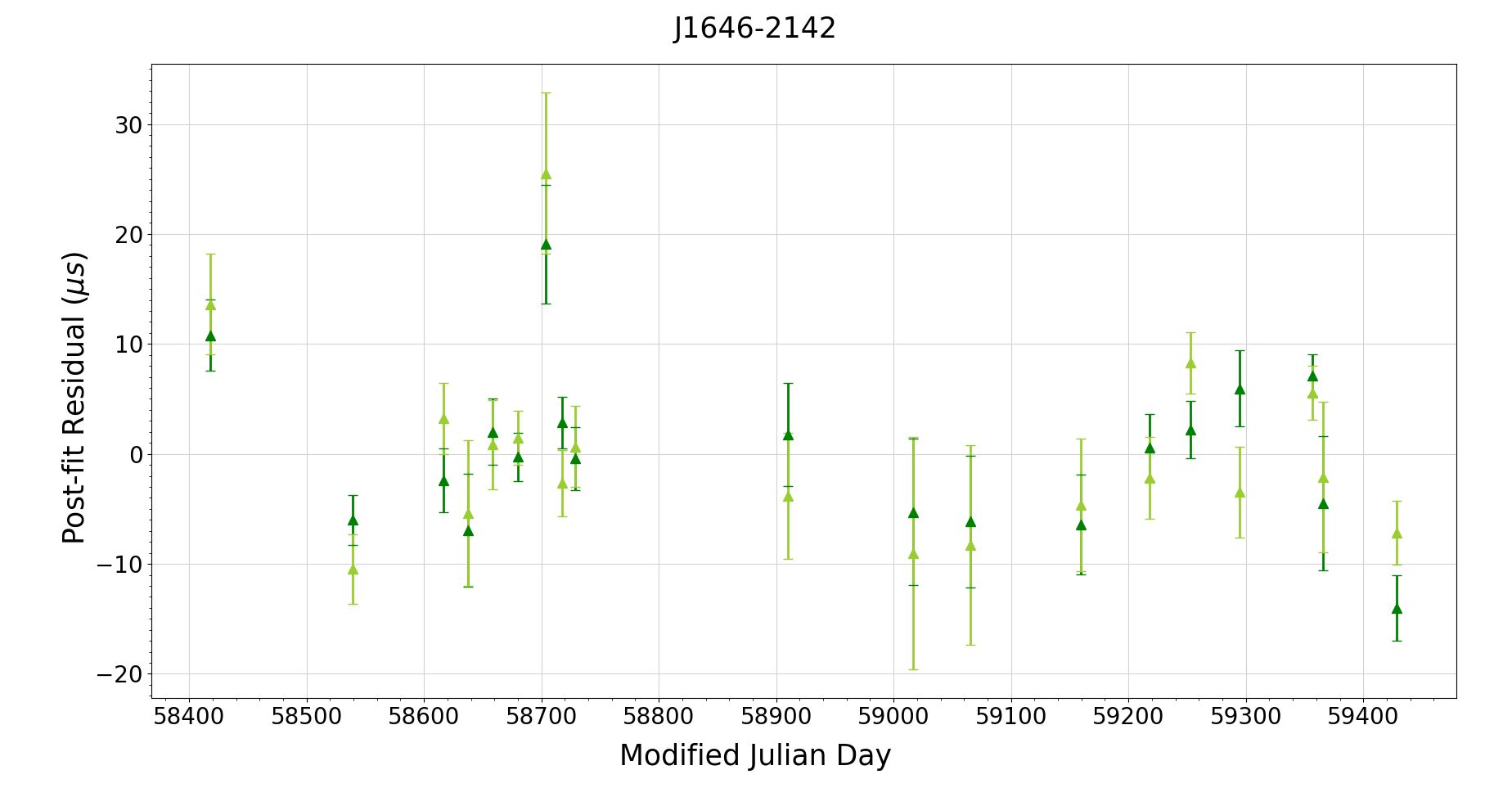}
                \vspace{-5mm}
    \caption{{Post-fit timing residuals versus MJD for the non-PTA pulsar J1646$-$2142. The colour and marker scheme are the same as Figure \ref{J1646_GMRT}.}}
    \label{J1646_GMRT_residuals}
\end{figure}

  \subsubsubsection{J1120$-$3618}
 {J1120$-$3618 is a 5.56 $ms$ pulsar in a binary system with 5.7 days of the orbital period.} For this pulsar, we see a single broad component with an unresolved feature near its peak in band-3 (Figure \ref{l}). {The W50 (width at the half of intensity peak) corresponding to its broad pulse component is  1.56 $\pm$ 0.04 $ms$.} Figure \ref{J1120_GMRT} and \ref{J1120_GMRT_residuals} show its DM variation with time and post-fit timing residual from NB and WB analysis.
\vspace{-3mm}
\subsubsubsection{J1646$-$2142}
{J1646$-$2142 is an isolated millisecond pulsar spinning with a period of 5.85 $ms$}. Out of four non-PTA pulsars, this MSP has interesting profile evolution with frequency (Figure \ref{j} and \ref{m}). {The W50 corresponding to its strongest pulse component in band-3 is  0.91 $\pm$ 0.05 $ms$. In band-4, the same component has a W50 of 0.69 $\pm$ 0.05 $ms$.} In band-3, the peak amplitude's ratio of the second to the first component changes from 0.29 to 0.55 from the {lowest} to {the highest-frequency subband.} In addition, the  separation between two components increases from 1.69(5) ms to 1.87(5) ms from the {lowest} to {the highest-frequency subband}, which is opposite to the radius-to-frequency mapping seen for some pulsars \citep{2012hpa..book.....L}. Similarly, the peak amplitude ratio of the two components increases from 0.96 to 1.32 from {the lowest} to {the highest-frequency subband} in band-4. However, the evolution of separation between the two peaks within band-4 is not significant and lies within $\pm$1 phase bin error. Figure \ref{J1646_GMRT} and \ref{J1646_GMRT_residuals} show DM variation with time and post-fit timing residual from NB and WB analysis for this pulsar.

\begin{figure}[H]
\centering
        \includegraphics[width=0.9\linewidth,keepaspectratio]{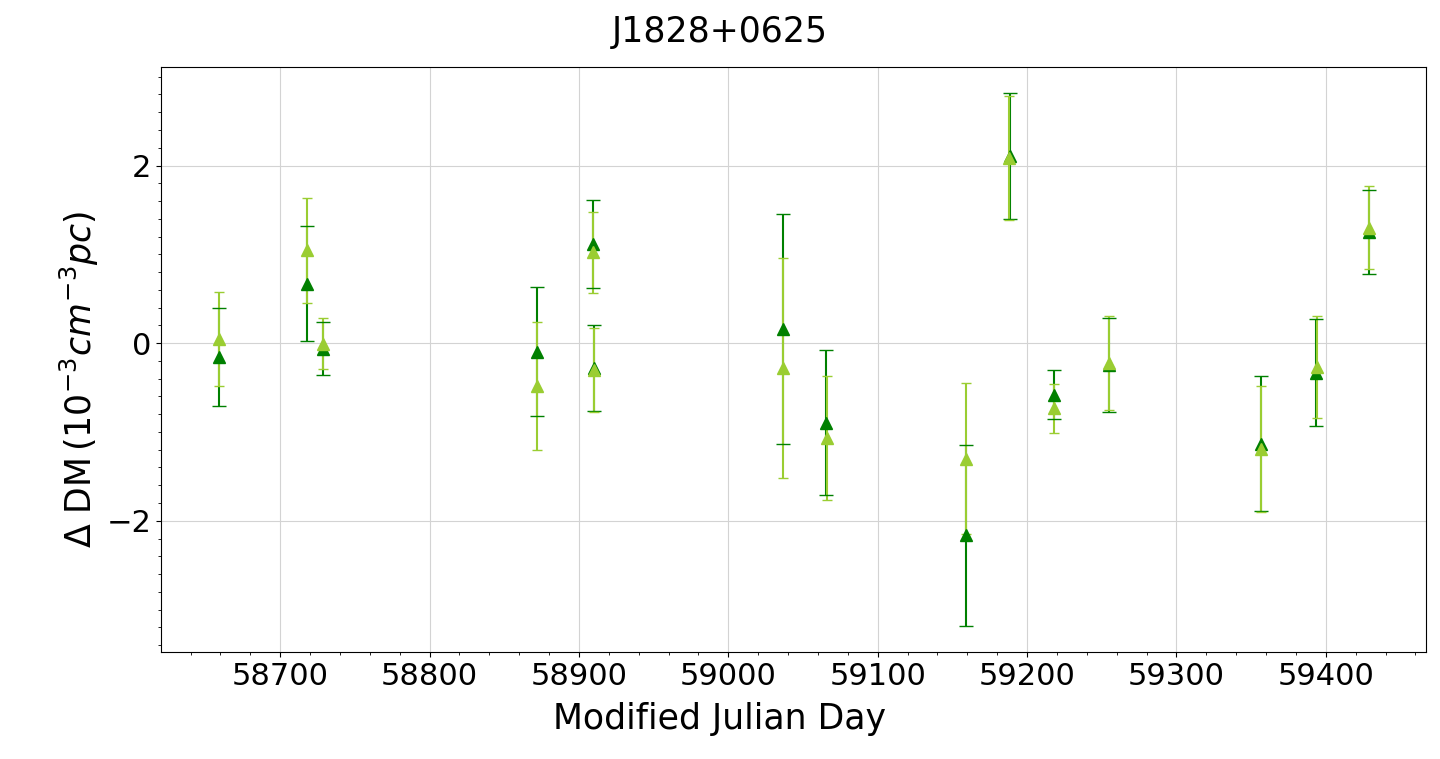}
        \vspace{-5mm}
    \caption{DM variation with time for the non-PTA pulsar J1828$+$0625 in band-3 of the uGMRT. {The colour and marker scheme are the same as Figure \ref{J1646_GMRT}.}}
    \label{J1828_GMRT}
\end{figure}

\begin{figure}[H]
\centering
       \includegraphics[width=0.9\linewidth,keepaspectratio]{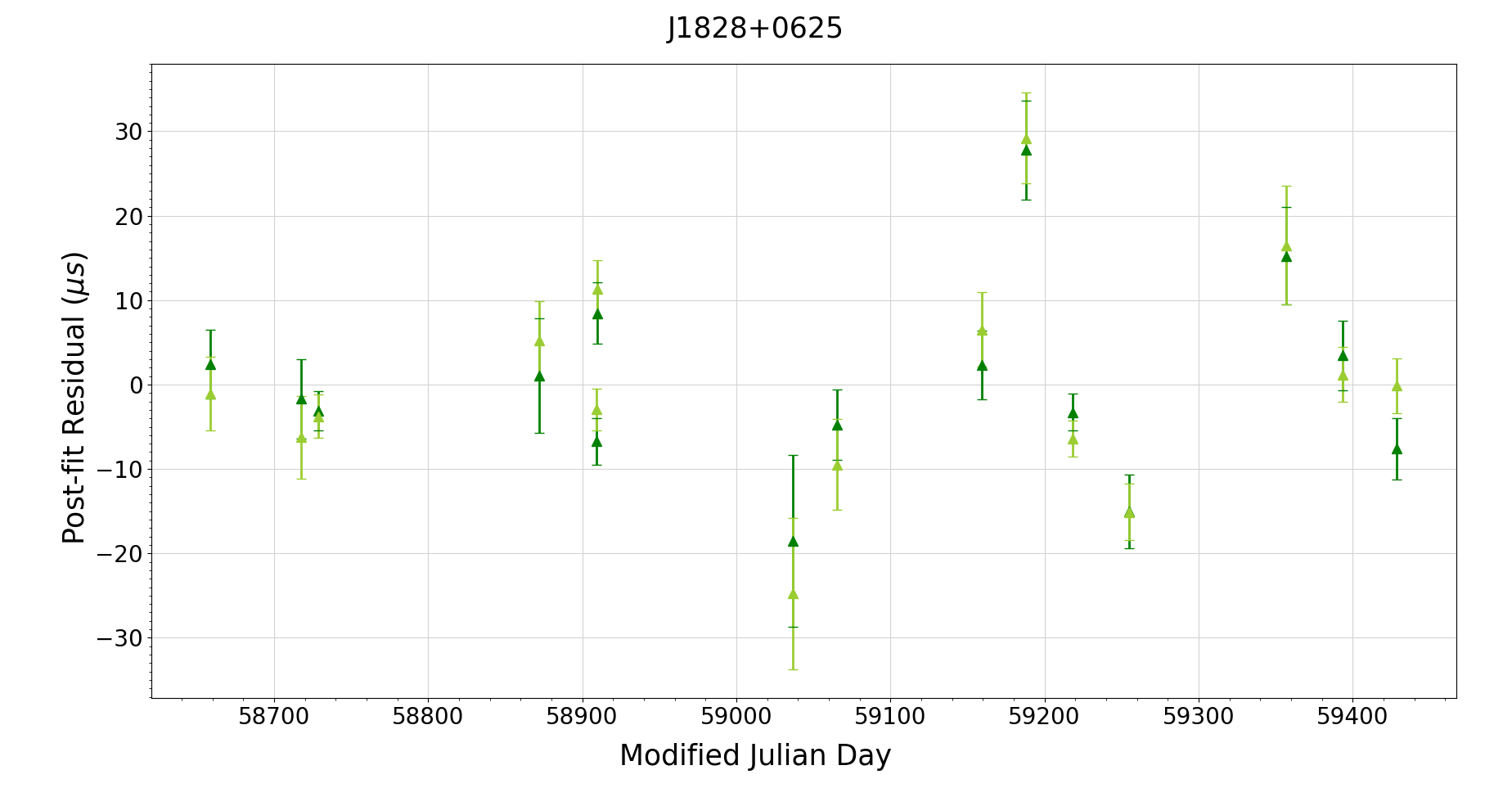}
        \vspace{-5mm}
    \caption{{Post-fit timing residuals versus MJD for the non-PTA pulsar J1828$+$0625. The colour and marker scheme are the same as Figure \ref{J1646_GMRT}.}}
    \label{J1828_GMRT_residuals}
\end{figure}

\begin{figure}[H]
\centering
        \includegraphics[width=0.9\linewidth,keepaspectratio]{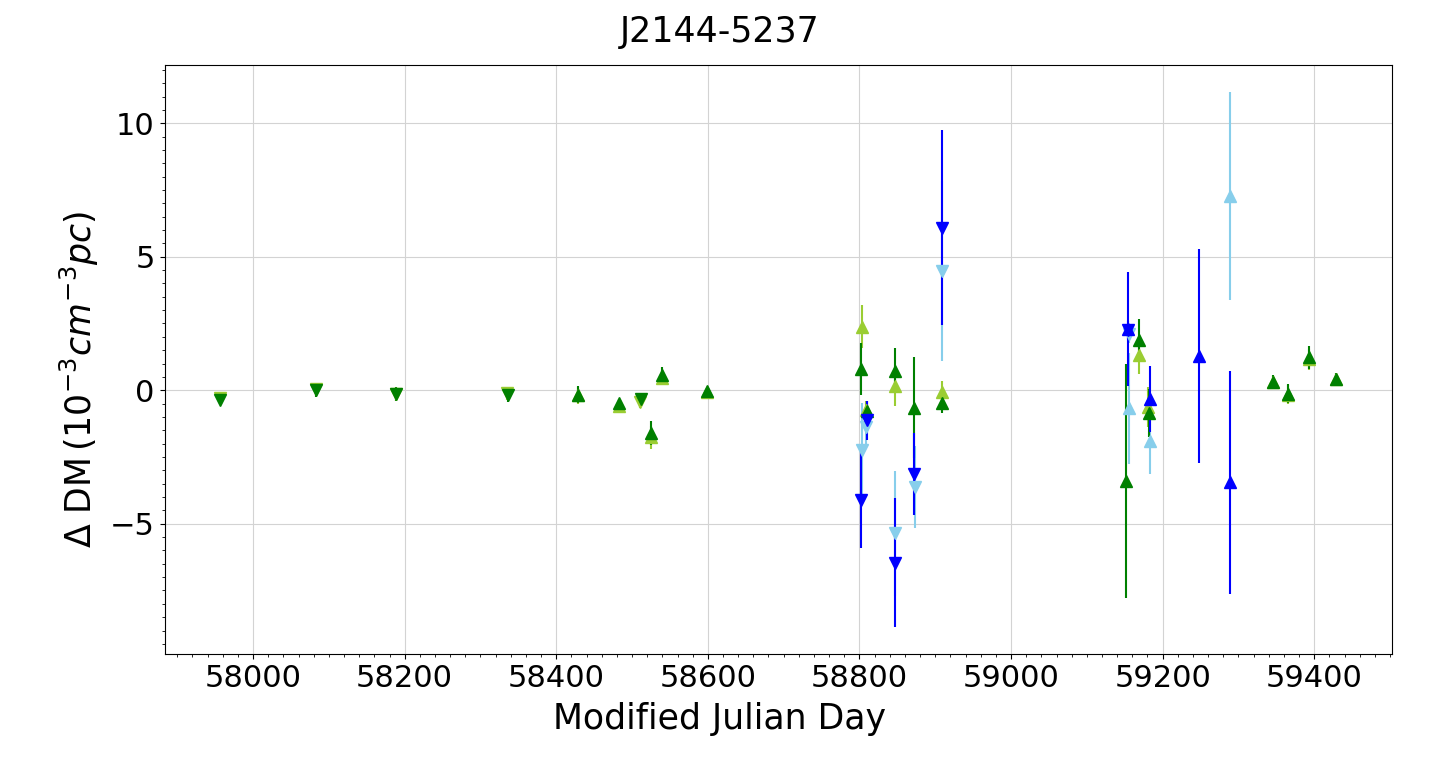}
        \vspace{-5mm}
    \caption{Figure showing DM variation with time for the non-PTA pulsar J2144$-$5237 in two separate bands of the uGMRT. The colour and symbol schemes are same as in Figure \ref{J1646_GMRT}.}
    \label{J2144_GMRT}
\end{figure}

\begin{figure}[H]
\centering
        \includegraphics[width=0.9\linewidth,keepaspectratio]{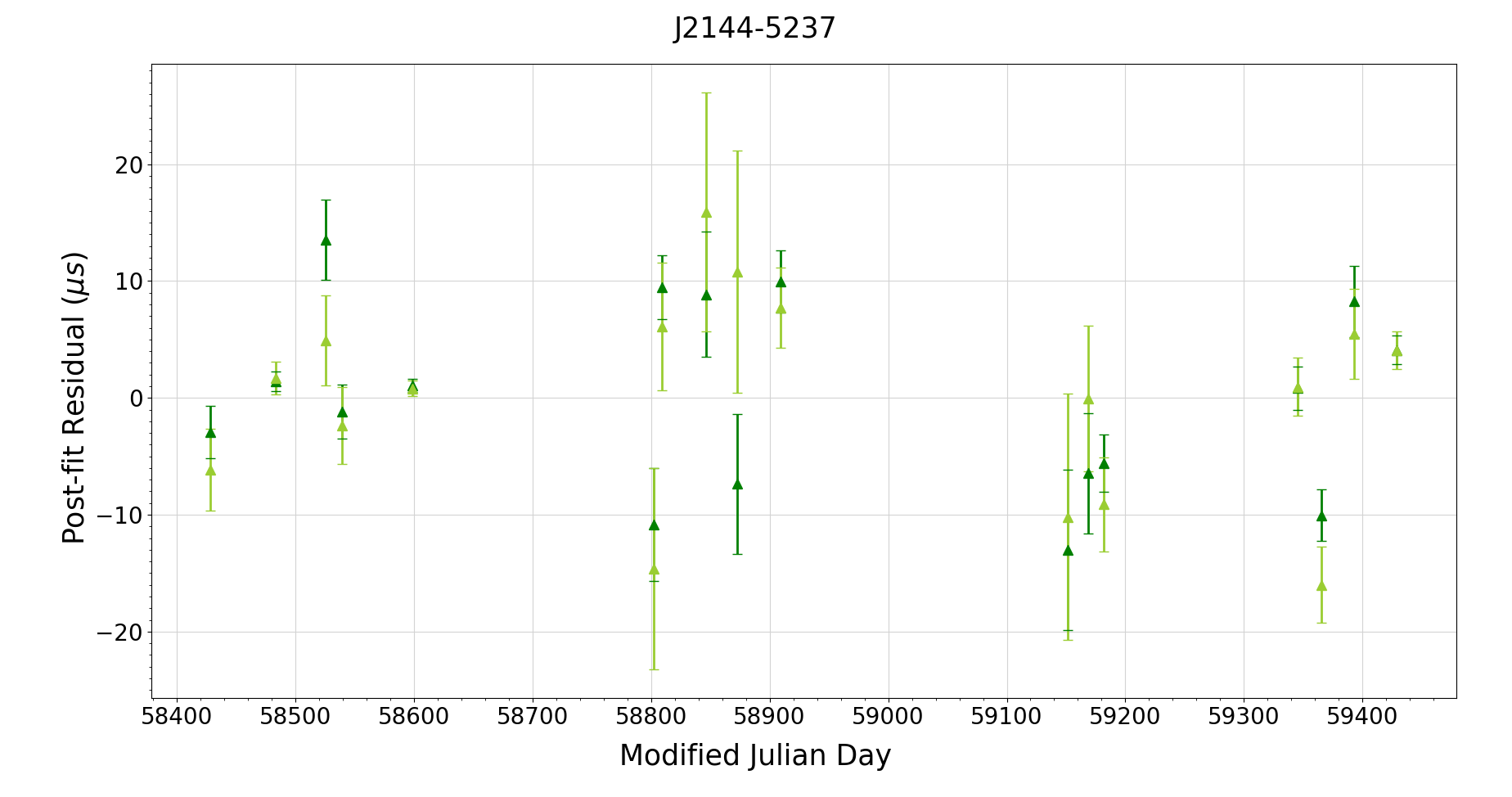}
        
        \vspace{-5mm}
    \caption{{Post-fit timing residuals versus MJD for the non-PTA pulsar J2144$-$5237. The colour and marker scheme are the same as Figure \ref{J1646_GMRT}.}}
    \label{J2144_GMRT_residuals}
\end{figure}

 \subsubsubsection{J1828$+$0625}
  {J1828$+$0625 is a 3.63 $ms$ pulsar in a binary system with 77.9 days of the orbital period.} It exhibits a single narrow component pulse profile in band-3 (Figure \ref{i}). {The W50 corresponding to its narrow pulse component is  0.57 $\pm$ 0.03 $ms$.} The temporal variation of DM and post-fit timing residual of this pulsars from NB and WB analysis are shown in Figure \ref{J1828_GMRT} and \ref{J1828_GMRT_residuals}, respectively.
\vspace{-5mm}
\subsubsubsection{J2144$-$5237}
 {J2144$-$5237 is a 5.04 $ms$ pulsar present in a binary system with 10.6 days of the orbital period.} {The W50 corresponding to its strongest pulse component (having two sub-components near its peak) in band-3 is 1.10 $\pm$ 0.04 $ms$. In band-4, the same component has a W50 of 1.02 $\pm$ 0.04 $ms$.} In band-3, the central component of J2144$-$5237 has two resolved peaks (Figure \ref{k}). In contrast to J1646$-$2142, the ratio of the second peak's amplitude to the first one decreases from 1.02 to 0.56 from {the lowest} to {the highest-frequency subband}. Unlike J1646$-$2142, J2144$-$5237 doesn't evolve much with frequency in band-4 (Figure \ref{n}). 
 Figure \ref{J2144_GMRT} and \ref{J2144_GMRT_residuals} show its DM variation with time and post-fit timing residual from NB and WB analysis. 

\subsubsection{Comparison between NB and WB timing for non-PTA pulsars}
The non-PTA pulsars show similar temporal DM variations for NB and WB analysis. No significant temporal variation in DM ($<$ $\pm$3$\sigma_{DM}$) is seen for the non-PTA pulsars except J1120$-$3618 (having $>$ $\pm$4$\sigma_{DM}$ variation of DM). Post-fit timing residual variations with MJD are also similar for NB and WB analysis for all the pulsars.

ToAs are more precisely estimated in WB than NB timing in all the cases. The median improvement in ToA uncertainty from NB to WB analysis is $\sim1.4$ times both in band-3 and band-4, respectively. In WB analysis, the median ToA uncertainty for non-PTA pulsars lies in the range of 2$-$5 $\mu s$ both in band-3 and band-4. The ToA uncertainties for J1646$-$2142 and J2144$-$5237 are almost the same in band-3 and band-4.

The median DM uncertainties are almost the same for NB and WB analysis both in band-3 and band-4. 
We achieve the median DM uncertainties in the order of 10$^{-4}$ and 10$^{-3}$ $\,pc~cm^{-3}$ in band-3 and band-4, respectively, using the WB timing. We find $\sim$8 and $\sim$4 times improvements in DM precision for J1646$-$2142 and J2144$-$5237, respectively, from band-4 to band-3.

 Number of eigenvectors ranges from 0-1 for non-PTA pulsars. The difference in median DM values from NB and WB analysis lies within $\pm1$ $\sigma_{DM}$. Thus median DM offsets between NB and WB timing are insignificant.
 We see negligible difference in ToA and DM uncertainties from coherently and incoherently dedispersed data sets of non-PTA pulsars. The increase of number of bins in coherently dedispersed profiles of non-PTA pulsars (requiring higher time resolution data) can possibly improve the DM and ToA precision.
 
\begin{figure}[H]
    \centering
    \includegraphics[width=0.8\linewidth, keepaspectratio]{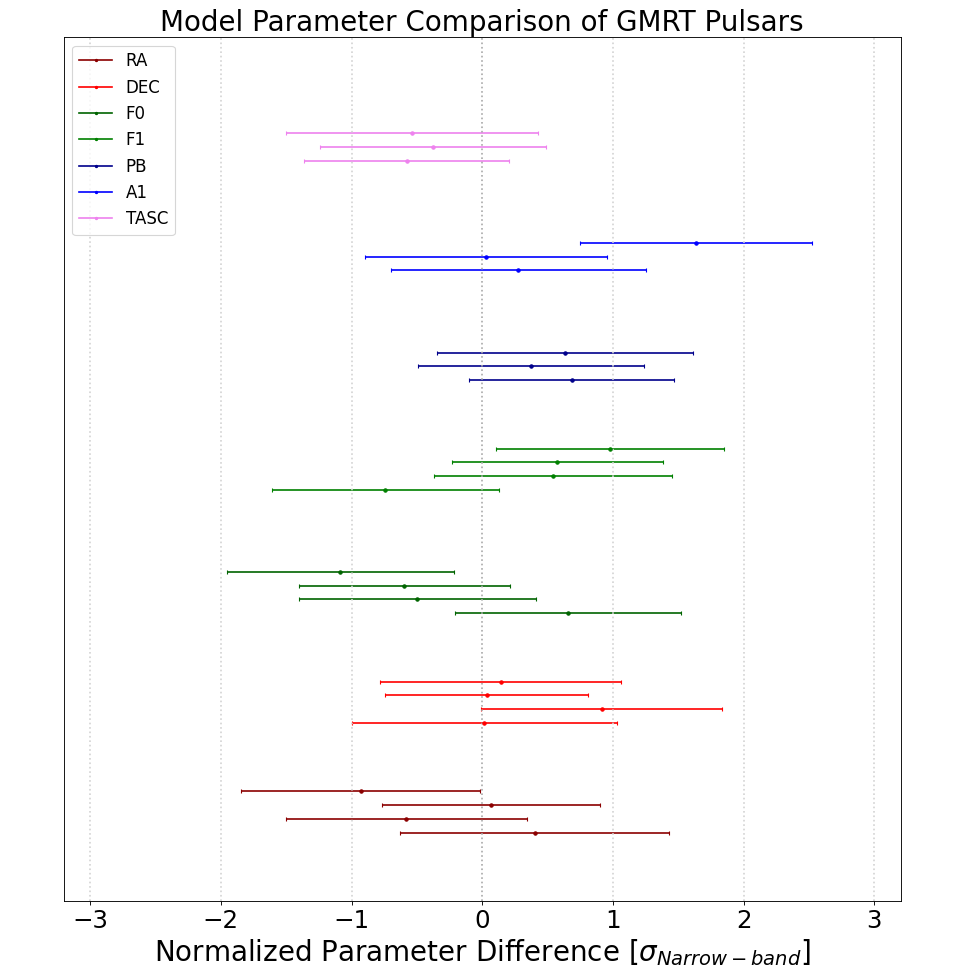}
    \caption{Figure showing a comparison {of our fitted} timing models of 4 non-PTA MSPs, in band-3, from NB and WB timing analysis. {Pulsars order (y-axis) for each parameter is maintained the same as in Table \ref{median_dm_table}.}The x-axis shows the differences of the fitted parameter (astrometric, spin, and binary) values normalized by the uncertainties from the NB timing model, i.e., $(X_{NB}-X_{WB})/\sigma_X^{NB}$ where X is the pulsar's model parameter. The error bars have a length equal to the ratio of parameter uncertainties from WB and NB models, i.e., $\sigma_X^{WB}/\sigma_X^{NB}$. Model parameters for all MSPs lie within $\pm 3 \sigma_{NB}$ for the two analyses.}
    \label{model_parameters}
\end{figure}

 From the timing fit, we achieve post-fit residual RMS (root-mean-square) of $<$ 10 $\mu$s for all the non-PTA pulsars from NB and WB analysis. The RMS of timing residuals is similar in NB and WB analysis. In the case of J1120$-$3618, we have accounted for the systematic increase in DM with time while fitting the model parameters. The precision of model parameters is similar in NB and WB timing. 
{In Fig. \ref{model_parameters} we compare the precision of fitted parameters of the timing models of the four non-PTA pulsars between NB and WB analysis}. We plot the differences of the fitted parameter values normalized by the uncertainties from the NB timing model ($\sigma_{NB}$). The error bars have a length of $\sigma_{WB}$/$\sigma_{NB}$. 
We find that the model difference of these two timing methods is well within $\pm$ 3$\,\sigma_{NB}$ confirming generation of similar long-term timing models from both the techniques (i.e. NB or WB).\\*

\begin{table}[H]
    
    \centering
    \begin{adjustbox}{width=0.7\columnwidth,height=8cm}
    \begin{tabular}{|c|c|c|c|}

 \hline
 
Parameter & PSR& NB Parameter& WB Parameter\\*
name (units)& & value & value\\*
 \hline
 \hline
RA & J1120-3618 & 11:20:23.350(1)& 11:20:23.350(1)\\
(hh:mm:ss.s) & J1646-2142 & 16:46:18.634(1)& 16:46:18.6347(9)\\
 & J1828+0625 & 18:28:28.9549(2)& 18:28:28.9549(2)\\
 & J2144-5237 & 21:44:35.6548(1)& 21:44:35.6550(1)\\
 \hline
DEC & J1120-3618 & -36:19:40.58(1)& -36:19:40.58(1)\\
(dd:mm:ss.s) & J1646-2142 & -21:42:02.5(1)& -21:42:02.4(1) \\
 & J1828+0625 & +06:25:09.808(7) & +06:25:09.808(5)\\
 & J2144-5237 & -52:37:07.522(4) & -52:37:07.521(4) \\
 \hline
 & J1120-3618 & 179.952669448(2) & 179.952669446(1)\\
F0 (s$^{-1}$) & J1646-2142 & 170.8494057187(3)& 170.8494057188(2)\\
 & J1828+0625 & 275.667196398(1)& 275.667196399(1)\\
 & J2144-5237 & 198.3554831469(2)& 198.3554831472(2)\\
 \hline
 & J1120-3618 & -7.4(7)& -6.8(6)\\
F1 (s$^{-2}$) & J1646-2142 & -2.40(1)& -2.41(1)\\
 & J1828+0625 & -3.6(1)& -3.64(8)\\
 & J2144-5237 & -3.57(2)& -3.58(2)\\
 \hline
 & J1120-3618 & 5.65994458(8)& 5.65994452(6)\\
PB (days) & J1828+0625 & 77.9249695(9)& 77.9249691(8)\\
 & J2144-5237 & 10.58031830(3)& 10.58031828(3)\\
 \hline
 & J1120-3618 & 4.30400(1)& 4.30399(1)\\
A1 (light-sec) & J1828+0625 & 34.888402(7)& 34.888402(6)\\
 & J2144-5237 & 6.361084(6)& 6.361075(5)\\
 \hline
 & J1120-3618 & 56225.01595(4)& 56225.01598(3)\\
TASC (MJD) & J1828+0625 & 57546.60298(2)& 57546.60299(2)\\
 & J2144-5237 & 57497.785575(4)& 57497.785577(4)\\
 \hline
 Post-fit& J1120$-$3618 & 6.9 & 6.3 \\
 timing& J1646$-$2142 & 6.1 & 5.0\\
 residuals&J1828$+$0625 & 9.1 & 7.7\\
 ($\mu s$)& J2144$-$5237& 3.8 & 3.8\\
 \hline
  & J1120$-$3618 & \multicolumn{2}{c|}{58538$-$59357 ; 58910}\\
  START - FINISH & J1646$-$2142 & \multicolumn{2}{c|}{58418$-$59429 ; 58910}\\
  ; Reference MJD& J1828$+$0625 & \multicolumn{2}{c|}{58659$-$59429 ; 59037}\\
 & J2144$-$5237 & \multicolumn{2}{c|}{58429$-$59429 ; 58909}\\
\hline
\end{tabular}
\end{adjustbox}
    \caption{The table lists parameter values and uncertainties obtained in NB and WB analysis. We have used only band-3 coherently dedispersed data in the timing model fit. {The planetary ephemeris DE200 is used for the barycentric correction of ToAs.} The first column contains parameter names fitted for a given pulsar (second column) and the units of their measurements. Fitted parameters: F0 - spin frequency of pulsar, F1 - time derivative of F0, RA - right ascension, DEC - declination; binary parameters: PB - binary orbital period, A1 - projected semi-major axis of the orbit, TASC - epoch of ascending node passage, {START-FINISH MJD - solution validity range, and Reference MJD - reference epoch for the measured parameters}. Binary parameters are not fitted for J1646$-$2142 as it is an isolated pulsar. {For the three binary pulsars, we have used the ELL1 binary model.}}
    \label{timing_solutions}
\end{table}

\section{Conclusions}
\label{sec:conclusions}

We provide a comparative study of the WB timing analysis with the conventional NB timing analysis at low frequency using the uGMRT for a set of non-PTA GMRT MSPs and for some well studied PTA MSPs. {ToAs are, in general, more precise in WB analysis than NB analysis. Though NB and WB timing provide similar DM precision for a given band. ToA precision is, in general, better in band-3 compared to band-5 for the PTA pulsars. For non-PTA pulsars, the ToA precision is similar in band-3 and band-4.} Also, band-3 of the GMRT provides much higher DM precision for all the eight MSPs compared to the other observing bands. 

{For PTA pulsars, we typically achieve sub-$\mu$s {ToA precision from WB analysis,} for {individual epochs both in band-3 and band-5}. For J1640$+$2224, J1909$-$3744 and J2145$-$0750 the DM precision obtained, in band-3, are in the order of 10$^{-5}$ $pc\,cm^{-3}$. For J1713$+$0747, similar DM precision can be achieved (following Eq. 3 of \cite{2020arXiv200908409J}) by combining near-simultaneous band-3 and band-5 observations. In band-5, the DM precision is of the order of $\sim$10$^{-3}$ $pc\,cm^{-3}$ for the PTA pulsars. The 
{median} DM values obtained from NB and WB timing for {PTA pulsars} are within the median DM uncertainties {except J1640$+$2224. Significant temporal variations of DM ($>\pm3\,\sigma_{DM}$) are observed for PTA pulsars.} From WB analysis the best ToA and DM precision we find in band-3 are $87\,ns$ and $1\times10^{-5}\,pc\,cm^{-3}$ for J2145$-$0750 and in band-5 are $278\,ns$ and $8\times10^{-4}\,pc\,cm^{-3}$ for J1909$-$3744. We have compared the ToA and DM precision for the commonly observed PTA pulsars with the earlier GMRT results in band-3 \citep{2021arXiv211206908N} and in band-5 \citep{2021arXiv210105334K} as well as with the inter-band DM measurements combining band-3 and band-5 \citep{2021arXiv210105334K}. 
The best median DM precision reported in this work is 3 times better in band-3 (for J2145$-$0750) and $\sim$ $10^2$ times better in band-5 (for J1909$-$3744) compared to earlier results. In addition, low-frequency intra-band DM estimates with the full GMRT array is more precised that the inter-band measurements using multiple sub-array.} 
{Thus the current work illustrates the maximum possible DM and ToA precision achievable for some of the best timed PTA pulsars with the wide-band system using the full timing sensitivity of the uGMRT.}

For non-PTA pulsars, the WB timing provides {$\mu s$ ToA precision} {both in band-3 and band-4}. The DM precision obtained are of the order of $\sim$10$^{-4}$ $\,pc~cm^{-3}$  and $\sim$10$^{-3}$ $\,pc~cm^{-3}$ in band-3 and band-4, respectively. The non-PTA pulsars {(having flux densities around 1-2 mJy at 400 MHz) are giving timing precision $< 10\mu s$ in NB and WB analysis at band-3.} The fitted model parameters from NB and WB analyses for the non-PTA pulsars in 2$-$4 years timing baseline agree well within $\pm$ 3 $\sigma$ uncertainties confirming the applicability of WB analysis for long-term timing. {In band-3, the timing precision is similar between NB and WB analysis for all four non-PTA pulsars. This work shows} the typical DM, {ToA}, and timing precision that can be achieved for newly discovered {GMRT} pulsars from low-frequency follow-up studies. {For non-PTA pulsars, the difference in median DM values from NB and WB analysis is less than the DM errors both in band-3 and band-4}. No significant temporal variations of DM ($<\pm3\,\sigma_{DM}$) are observed for the GMRT pulsars, except J1120$-$3618.

 In the case of the non-PTA pulsars even with an order of magnitude lower flux densities than the PTA MSPs, the achieved DM precision (Table \ref{Table_2}), is comparable with the higher frequency measurements for PTA MSPs \citep{2021ApJS..252....5A} making them as potential candidates to include in the IPTA experiment in a search for a GW background aided by the more sensitive upcoming telescopes providing better ToA precision. Since at the intermediate signal regime of stochastic background of GWs, the detection significance strongly depends on the number of pulsars included in the array 
 \citep{2013CQGra..30v4015S}, such low frequency follow-up aided with WB timing can play an important role.
Following the work by \cite{2021arXiv211206908N} for PTA MSPs and the current work for newly discovered MSPs, the prospect of using a low frequency observing facility at a sensitive telescope like the uGMRT for high precision timing studies to aid the global PTA efforts is clearly evident. 

Moreover, the timing with the full GMRT array (70\%) in band-3 (with a gain of 7 K/Jy), presented here, is complementary to the currently existing WB timing facilities like MeerKAT and CHIME provoding lowest frequency coverage of 580$-$1670 MHz \citep{2020PASA...37...28B} and 400$-$800 MHz \citep{2021ApJS..255....5C}, respectively.

\section{Acknowledgments}
We acknowledge the support of the Department of Atomic Energy, Government of India,
under project no. 12-R$\&$D-TFR-5.02-0700. The GMRT is run by the institute National Centre for
Radio Astrophysics of the Tata Institute of Fundamental Research, India. We thank the anonymous referee for comments that improved the quality of the paper. We acknowledge the support of GMRT telescope operators for observations.

\end{document}